\documentclass[]{spie}  

\usepackage{amsmath,amsfonts,amssymb}
\usepackage{graphicx}
\usepackage[colorlinks=true, allcolors=blue]{hyperref}

\newcommand\tierras{\textit{Tierras} }
\newcommand\arcdeg{\mbox{$^\circ$}}%
\newcommand\arcmin{\mbox{$^\prime$}}%
\newcommand\arcsec{\mbox{$^{\prime\prime}$}}%

\graphicspath{{./}{figures/}}
\usepackage{deluxetable}
\usepackage{commath}

\title{The {\em \LARGE \bfseries Tierras} Observatory: An ultra-precise photometer to characterize nearby terrestrial exoplanets}

\author[a]{Juliana García-Mejía}
\author[a]{David Charbonneau}
\author[a]{Daniel Fabricant}
\author[a]{Jonathan M. Irwin}
\author[a]{Robert Fata}
\author[a]{Joseph M. Zajac}
\author[a]{Peter E. Doherty}
\affil[a]{Center for Astrophysics $|$ Harvard and Smithsonian, 60 Garden Street, Cambridge MA}

\authorinfo{Corresponding author: Juliana García-Mejía, \\ juliana.garcia-mejia@cfa.harvard.edu}

\begin{document} 
\maketitle

\begin{abstract}
We report on the status of the \tierras Observatory, a refurbished 1.3-m ultra-precise fully-automated photometer located at the F. L. Whipple Observatory atop Mt. Hopkins, Arizona. \tierras is designed to limit systematic errors, notably precipitable water vapor (PWV), to 250 ppm, enabling the characterization of terrestrial planet transits orbiting $< 0.3 \, R_{\odot}$ stars, as well as the potential discovery of exo-moons and exo-rings. The design choices that will enable our science goals include: a four-lens focal reducer and field-flattener to increase the field-of-view of the telescope from a $11.94 \arcmin$ to a $0.48 \arcdeg$ side; a custom narrow bandpass ($40.2$ nm FWHM) filter centered around $863.5$ nm to minimize PWV errors known to limit ground-based photometry of red dwarfs; and a deep-depletion $4K \times 4K$ CCD with a 300ke- full well and QE$>85\%$ in our bandpass, operating in frame transfer mode. We are also pursuing the design of a set of baffles to minimize the significant amount of scattered light reaching the image plane. \tierras will begin science operations in early 2021.


\end{abstract}


\section{Introduction} \label{sec:intro} 

The construction of observatories tailor-built to routinely achieve precise photometry from the ground is motivated by the multitude of exoplanetary and stellar phenomena whose exploration will be enabled with high-cadence, ultra-precise time series of diverse stellar targets. One such exciting phenomenon concerns the transit detection and characterization of terrestrial, or rocky, planets, with particular attention towards $< 1.5 \, R_{\oplus}$ planets. For at least the next 15 years, the only spectroscopically accessible terrestrial exoplanets will be those that orbit nearby mid-to-late M-dwarfs, with radii less than $0.3 \,R_{\odot}$. Owing to these hosts' small sizes and low luminosities, the atmospheres of rocky planets around them can be probed with the methods of transmission and emission spectroscopy using the upcoming James Webb Space Telescope \cite{Mo17}, and ground-based Extremely Large Telescopes (ELTs) \cite{Sn13}. Although six systems of transiting terrestrial planets have been found orbiting such stars within 15 parsecs (GJ1132\cite{Be15}, LHS 1140\cite{Di17}, TRAPPIST-1\cite{Gi16}, LHS 3844\cite{Va19}, LTT 1445\cite{Wi19}, and TOI-540\cite{Men20}), we may still be missing temperate rocky worlds analogous to the smallest Solar System planets (with radii below $1 \, R_{\oplus}$). Identifying a $1 \, R_{\oplus}$ planet around a $0.1 \, R_{\odot}$ (M7), $0.2 \, R_{\odot}$ (M5), and $0.3 \, R_{\odot}$ (M3) star with $3\sigma$ significance requires a per-point photometric precision ($(1/3)(R_{p}/R_{*})^2$) of 2795, 700, and 311 ppm, respectively. A Venus-sized ($0.95 \, R_{\oplus}$) planet around the same hosts would require 2522, 631, and 280 ppm, respectively. 
 
Other planetary phenomena that could be unlocked with ultra-precise photometric observations include the hitherto elusive discoveries of exo-moons and exo-rings \cite{Sa99}. By analogy to their abundant Solar System counter-parts, exo-satellites may serve as tracers of the planet formation environments and migration histories of their hosts (See Ref. \citenum{He18} for a review). For example, Refs. \citenum{Po74} and \citenum{Ca02} placed constraints on the temperature distribution of the accretion disk from which Jupiter's Galilean moons likely formed based on the satellites' compositions, and Ref. \citenum{Sa06} explored the relationship between the size of the particles that make up Saturn's rings and the geophysical and tidal activity around the planet. Despite numerous searches for exo-satellites (e.g., Refs. \citenum{Ch06, Ki12}), no definitive detections have been announced to-date. The photometric precision requirements to detect exo-moons and exo-rings are achievable for the largest of these phenomena. For instance, the transit detection of a Ganymede-sized ($0.41 \, R_{\oplus}$) satellite around a $0.1 \, R_{\odot}$ (M7), $0.2 \, R_{\odot}$ (M5), and $0.3 \, R_{\odot}$ (M3) star with $3\sigma$ significance requires an observatory capable of delivering time-series with photometric errors below 477, 119, and 53 ppm, respectively. In the case of exo-rings, projection and forward-scattering effects complicate the assessment of the photometric precision required to detect them \cite{Zu15,Ba04}. However, Ref. \citenum{Ba04} argued that a photometric precision between 100 and 300 ppm with 15 min time resolution, accompanied by  a careful fitting effort, would be sufficient to detect Saturn-like ring systems.  

On the stellar side, ultra-precise time series enable the precise measurements of stellar rotational modulations for nearby mid-to-late M dwarfs, with significant gains for the faintest of these targets. Stellar rotation is a directly observable probe that provides insights into a star's magnetic field and dynamo generation (e.g., Ref. \citenum{Wr11}), and as such is likely linked to a slew of stellar activity, including flares, spots, coronal and chromospheric emission (e.g., Refs. \citenum{Sk72, Ma89}). Substantial and recent observational progress has been made in measuring stellar rotation periods for the smallest stars. For example, Refs. \citenum{Ne16} and \citenum{Ne18} measured the rotation periods of 387 (Northern) and 234 (Southern) mid-to-late M-dwarf stars, respectively. They found that the measurement of rotation periods becomes more challenging for the later spectral types, as photon noise and precipitable water vapor errors limit the relative flux changes that can be measured over these stars' rotational timescales. 

State-of-the-art facilities dedicated to the study of mid-to-late M dwarfs, such as MEarth \cite{Nu08, Ir15}, TRAPPIST\cite{Gi11, Je11}, and SPECULOOS-South\cite{Mu20}, achieve photometric precisions of a few mmag: MEarth can reach per-point precisions between 2 and 8 mmag for stars with radii between $0.3$ and $0.15 \, R_{\odot}$ after correcting for systematics and stellar activity \cite{Be12}; TRAPPIST generally reports 2 minute photometric precision errors in the order of $0.2-0.5\%$ for late M dwarfs (e.g., Refs. \citenum{Gi12, Gi16}); and SPECULOOS-South recently demonstrated the ability to reach a median 30-min bin precision of 1.5 mmag in their raw, non-detrended light curves of ultracool dwarfs (spectral type M7 and later) \cite{Mu20}. The photometric precision of these and other ground-based observatories is ultimately limited by photon noise and a combination of atmospheric effects, with scintillation and differential extinction being particularly pernicious (Section \ref{subsec:precision}). 

A dedicated space-based photometric observatory that altogether avoids observing through the Earth's atmosphere would seem ideal to address ground-based limitations. NASA's Transiting Exoplanet Survey Satellite (TESS) (launched 2018 April) is one such example \cite{Ri15}. TESS is a two+ year (nearly) all-sky survey producing high cadence photometric time series for at least 200,000 of the closest stars. For the M-dwarf candidates, TESS data will suffer from two primary limitations. First, TESS light curves of the relatively faint M-dwarfs will be photon-noise limited owing to its 10 cm aperture. Second, TESS will often miss temperate rocky planet, exo-moon, and exo-ring candidates at longer orbital periods due to its limited baseline. 

Motivated by this context, we are pursuing the construction of The \tierras\footnote{\tierras is the plural of `Earth', in Spanish.} Observatory, a fully-automated ultra-precise ground-based facility. \tierras is designed to limit systematic errors, notably precipitable water vapor (PWV), to 250 ppm, and stochastic errors, notably scintillation and target photon noise, to less than that of an Earth-transit for an integration time of 5 minutes, for all stars smaller than $0.3 R_{\odot}$ out to a distance of 15 pc. Such a precision will enable the discovery of terrestrial planets, exo-moons, exo-rings, the measurement of stellar rotational modulations, and the follow-up of exciting TESS discoveries. \tierras will repurpose the dormant 1.3-m telescope atop Mount Hopkins, Arizona, which hosted the 2MASS North IR camera almost twenty years ago \cite{Sk06}. Following a brief discussion of the main challenges associated with reaching high photometric precision from the ground in Section \ref{subsec:precision}, we outline our ongoing efforts to refurbish and transform the telescope into an ultra-precise photometer in Section \ref{subsec:overview}. 

\subsection{Photometric Precision Limits from the Ground} \label{subsec:precision}

Adapting the formalisms outlined in Refs. \citenum{Nu08, Co17}, and \citenum{St17}, we quantify the total root-mean-square (RMS) photometric error budget (i.e., the achievable photometric precision) of a ground-based observatory, $\sigma_{\rm tot}$, via the following equation:
\begin{equation}\label{eq:toterr}
    \sigma_{\rm{tot}} = \sqrt{\sigma_{\rm{rel \, flux}}^2 + \sigma_{\rm{scint}}^2 + \sigma_{\rm PWV}^2}
\end{equation}

where $\sigma_{\rm rel \, flux}$ is the normalized relative flux error associated with counting photons via a CCD detector,  $\sigma_{\rm scint}$ is the scintillation error for an ensemble of a target and uncorrelated reference stars, and $\sigma_{\rm PWV}$ accounts for the atmospheric variability error associated with the amount of PWV present in the atmospheric column above the observatory. We describe these three terms in detail below. 

The first term in Eq. (\ref{eq:toterr}), $\sigma_{\rm rel \, flux}$, is associated to the total noise for a CCD aperture photometry measurement due to the target and reference stars, and accounts for the photon noise due to the target and reference stars, the noise contribution from sky background pixels, the number of dark current electrons per pixel, and the RMS readout noise (e-/pixel) \cite{Me95, Co17}. For our application, neither detector electronic noise nor sky photon noise are important contributions. Assuming that there exists a sufficient number of bright nearby reference stars, the normalized relative flux error can be quantified simply as $\sigma_{\rm rel \, flux} = 1/\sqrt{N_{*}}$, where $N_{*}$ is the number of detected target photons. 

The second term in Eq. (\ref{eq:toterr}), $\sigma_{\rm scint}$, represents scintillation, which refers to the twinkling (angular excursions, distortions, and brightness fluctuations) of stars caused by turbulence in the Earth's high-altitude atmospheric layers \cite{Yo67}. Ref. \citenum{Ko12} showed that the strength of scintillation in the context of differential photometry depends on the amount of uncorrelated comparison stars, with correlated targets being $< 20 \arcsec$ apart. Assuming an uncorrelated comparison star ensemble, Ref. \citenum{St17} defines
\begin{equation}\label{eq:scinterr}
    \sigma_{\rm{scint}} = 1.5 \sigma_{s} \sqrt{1+1/n_E}
\end{equation}
where $\sigma_{s} = 0.09D^{-2/3} \chi^{7/4} (2t)^{-1/2} {e}^{(-h/h_o)}$ is the expected scintillation noise for a single star \cite{Yo67}. Here, $D$ is the telescope diameter (cm), $\chi$ is the airmass, $t$ is the exposure time (sec), $h$ is the observatory altitude (m), and $h_o = 8000$ m is the atmospheric scale height. The constant $0.09$ has units of cm$^{2/3}$s$^{1/2}$. 

The third term in Eq. (\ref{eq:toterr}), $\sigma_{\rm PWV}$, accounts for differential extinction, a noise source brought about by the time-dependent variation of molecular absorption in our atmosphere. Differential extinction of the first order is caused by the change in airmass throughout an observing night, and as such is a systematic effect that can generally be detrended with great success (e.g., Ref. \citenum{Ma11}). Differential extinction of the second order refers to the color-dependent difference in the amount of molecular absorption incurred by red target stars and blue ensemble stars, particularly due to the variable amount of precipitable water vapor (PWV) in the vertical column above the observatory. Previous efforts (e.g., Refs. \citenum{Bl08,Be12}) have shown that ground-based photometry of red dwarfs is primarily limited by this second-order differential extinction effect. 

We quantify PWV error, $\sigma_{\rm PWV}$, as the change in the apparent magnitude difference between a target star and a comparison star, as observed at two extremes of the PWV range ($0$ and $12$ mm of precipitable water vapor in a column above the observatory). As a ratio of fluxes, this corresponds to  
\begin{align}\label{eq:pwverr}
    \sigma_{{\rm PWV}} = \abs{1 - \frac{(F_1/F_2)_{\rm PWV = 0mm}}{(F_1/F_2)_{\rm PWV = 12mm}}}
\end{align}
where $F_{1}$ and $F_{2}$ are the flux (in ergs s$^{-1}$ cm$^{-2}$) of the target star and comparison star through our system, respectively. 


Additional noise sources may impact the ground-based photometric error budget, including the motion of the star on the detector and imperfect knowledge of the flat fielding correction. However, these are correctable with improved knowledge of the flat-field. 

\subsection{Addressing Photometric Precision Limits with \tierras} \label{subsec:overview}

\tierras is designed to limit the PWV error contribution to the total photometric noise to 250 ppm, and stochastic errors, notably scintillation and target photon noise, to less than that of an Earth-transit for an integration time of 5 minutes, for all stars $< 0.3 R_{\odot}$ out to a distance of 15 pc. This technical requirement translates into a series of design choices to overcome the factors that limit photometric precision described in Sec. \ref{subsec:precision}. \emph{First}, we must simultaneously monitor a large number of field stars, which serve as atmospheric calibrators; this in turn requires that we build a four-lens focal reducer and field-flattener to increase the field-of-view (FOV) of the 1.3-m telescope from a modest $11.94 \arcmin \times 11.94 \arcmin$ square to a $0.56 \arcdeg$ diagonal (Section \ref{sec:optics}). \emph{Second}, we intend to desensitize our observations to second-order differential extinction effects by using a custom narrow (40.2 nm FWHM) bandpass filter centered at $863.5$ nm. This bandpass avoids water vapor telluric features altogether, and encompasses a portion of the stellar spectrum where cool stars emit copiously (Section \ref{sec:filter}). \emph{Third}, we will use a $4K \times 4K$ deep-depletion CCD in fast frame transfer (shutter-less) mode. Our chip has a deep full well (300 ke-, compared to 100ke- for non-deep-depletion CCDs), high quantum efficiency ($>85\%$ in our bandpass, compared to $35\%$ for non-deep-depletion CCDs), low read-noise and low dark current (Section \ref{sec:ccd}). \emph{Fourth}, \tierras must be fully robotic in order to maximize our time on sky monitoring interesting targets. We have completed an extensive refurbishment and Telescope Control System (TCS) upgrade of the 1.3-m telescope, and will build upon the robotization efforts of the MEarth Observatory\cite{Ir15} to efficiently automate \tierras (Section \ref{sec:refurb}). The expected system performance is summarized in Section \ref{sec:perf}. \emph{Finally}, we discuss the status of the \tierras baffling system, crucial towards minimizing scattered light at the image plane, and the potential future installation of a nano-fabricated beam-shaping diffuser\cite{Ra04} to mold the point spread function as a broad and stable top-hat (Section \ref{subsec:future}).

\section{The \tierras Focal Reducer and Field Flattener} \label{sec:optics}

As discussed in Section \ref{subsec:precision}, ground-based photometric precision is hindered by the variable transmission of the atmosphere. Precisely quantifying a target star's brightness relies on monitoring other nearby stars to correct for changes in brightness arising from variations in atmospheric transparency. Distilled into a technical requirement, increasing the amount of comparison stars in the target's field is straightforwardly achieved by maximizing the observatory's effective FOV. In its original optical configuration, the 1.3-m telescope would only provide a modest $11.94 \arcmin \times 11.94 \arcmin$ FOV, assuming that we placed a $4K \times 4K$ ($15 \mu$m pixel size, $60$ mm side) detector at the Cassegrain focus. Thus, we set out to design a focal reducer and field flattener to maximize the FOV with minimal optical complexity; i.e., to get as close as possible to observing a $0.50 \arcdeg$ square of sky \cite{Nu08}. Aiming for a larger FOV was impractical due to vignetting at the primary mirror hole ($D_{\rm hole} = 200$ mm). We ultimately converged on a four-lens focal reducer and field flattener design that enables us to fit a $0.48 \arcdeg$ square of sky within the entire imaging area of the new \tierras detector (Section \ref{sec:ccd}). We describe the design and fabrication process for these auxiliary optics below, after a quick overview of the 1.3-m telescope's history and optical specifications.  

\subsection{A Brief History of the 1.3-m Telescope}\label{subsec:history}

\begin{figure} [ht]
   \begin{center}
   \begin{tabular}{c} 
   \includegraphics[height=6cm]{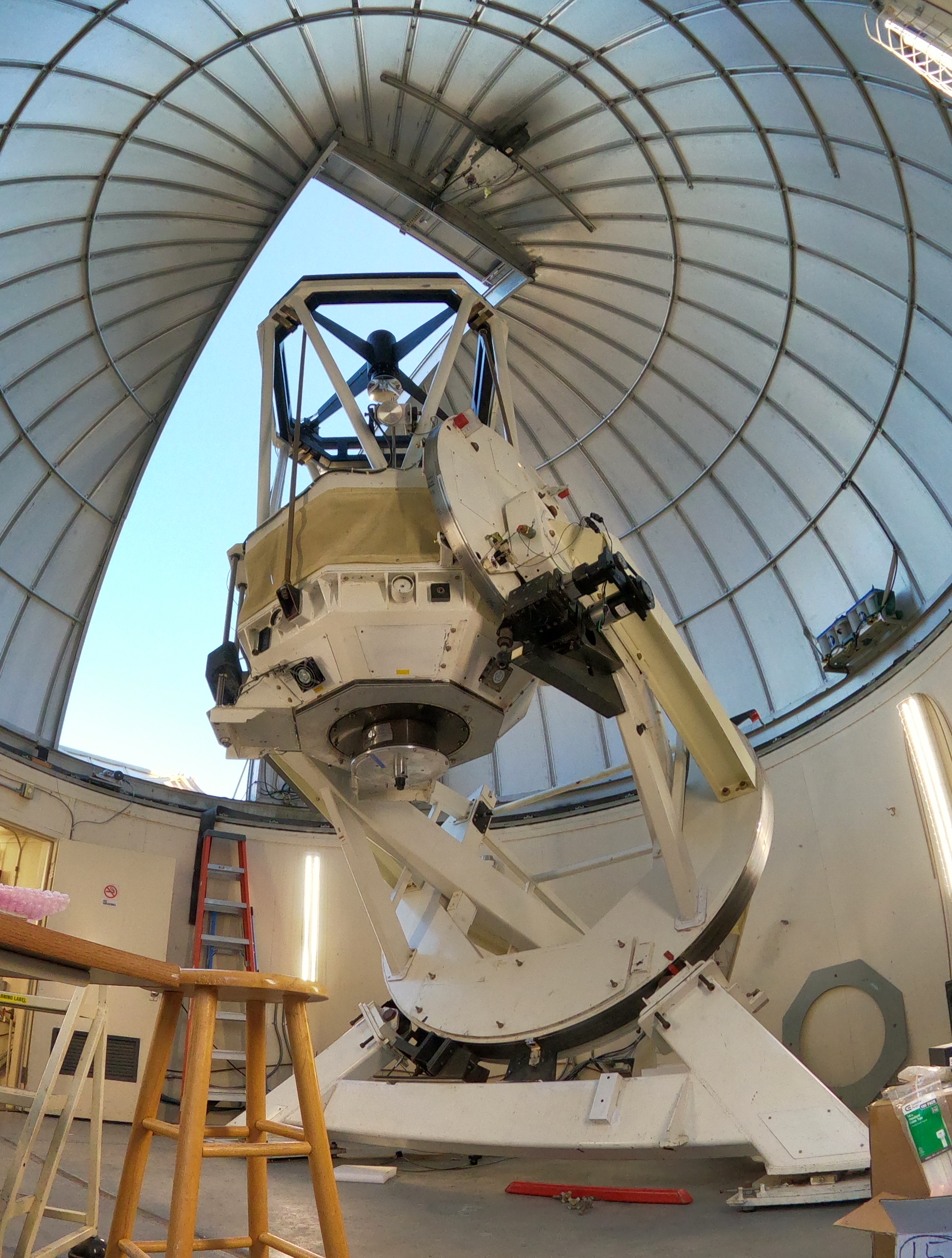}
   \includegraphics[height=6cm]{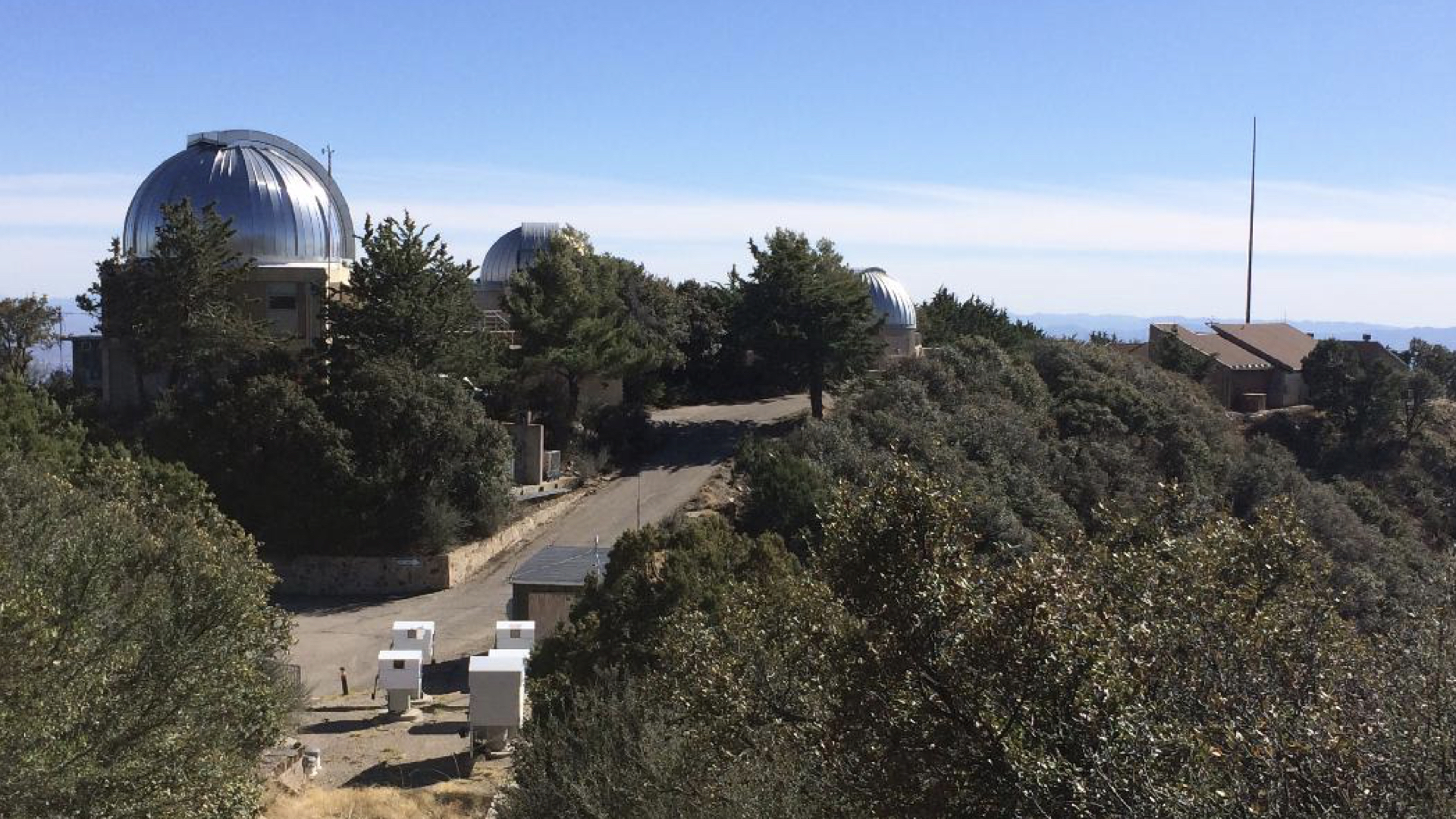}
   \end{tabular}
   \end{center}
   \caption[example] {\label{fig:1.3m} \footnotesize
 \emph{Left:} The 1.3-m telescope pictured inside its dome during the refurbishment process, which took place 2020 January (Section \ref{sec:refurb}). \emph{Right:} The FLWO ridge. Starting from the top left of the image (moving clockwise): the Tilinghast Reflector 1.5-m telescope, which hosts the TRES Spectrograph; the 1.2-m telescope, which hosts \emph{KeplerCam}; our 1.3-m telescope; The MINiature Exoplanet Radial Velocity Array (MINERVA); the Hungarian-made Automated Telescope (HAT-5). The MEarth Observatory (not pictured) is located near the ridge.}
   \end{figure}

\tierras utilizes a refurbished 1.3-m Cassegrain equatorial telescope, built to carry out the northern portion of the near-infrared (NIR) Two Micron All Sky Survey (2MASS) \cite{Sk06} (Figure \ref{fig:1.3m}). The telescope is part of the Fred Lawrence Whipple Observatory (FLWO), and sits at an elevation of 2306 m on a ridge below the summit of Mt. Hopkins, Arizona. The observatory coordinates are N $31\arcdeg 40\arcmin 59.61\arcsec$, W $110\arcdeg 52 \arcmin 31.14\arcsec$. The parabolic primary mirror has a radius of curvature of $R=5200$ mm, and a clear diameter of $D_{\rm 1, clear} = 1280$ mm. The hyperbolic ($K = -1.847$) secondary mirror has $R = 965.7$ mm, and $D_{\rm 2, clear} = 228$ mm. Both mirrors are made of Corning ultra-low expansion glass. They provide a Cassegrain focal ratio of $f/13.5$ and a plate scale of $11.94 \arcsec$ mm$^{-1}$ at the native focal plane, located $527$ mm behind the primary vertex. Rayleigh Optical Corporation polished the two mirrors, while M3 Engineering and Technology Corp built the telescope between 1995 and 1996, and commissioned it in 1997. 

Between 1997 June and 2001 February, the 2MASS North camera scanned $92\%$ of the northern hemisphere using three $256 \times 256$ NICMOS3 (HgCdTe) arrays and a couple of dichroic mirrors to enable simultaneous imaging of an $8.5\arcmin \times 8.5\arcmin$ FOV in the NIR $J$($1.25 \, \mu$m), $H$($1.65 \, \mu$m), and $K_s$($2.16 \, \mu$m) bands \cite{Mi96}. In lieu of a wide-field survey, 2MASS opted for an observation strategy consisting of steadily scanning the sky in declination at a rate of $57 ''$ s$^{-1}$, for an effective FOV per 7.8 sec exposure of $8.5\arcmin \times 6 \arcdeg$ \cite{Sk06}. Between 2003 and 2013, the telescope and the 2MASS South camera were repurposed to create The Peters Automated Infrared Imaging Telescope (PAIRITEL), the first meter class system designed to detect gamma ray burst transient events in the NIR fully robotically\cite{Bl06}. The telescope has been unused since 2013.

\subsection{Optical Design}\label{subsec:optdesign}

In its original optical configuration, the 1.3-m telescope is a very slow system ($f/\# = \rm{EFFL} / {D_{\rm 1, clear}} = 13.5$, where EFFL is the effective focal length), which translates into a narrow telescope plate scale ($S$) at Cassegrain focus, and therefore a small FOV. Reaching the desired $0.50 \arcdeg $ square FOV requires $f/\# \leq 5.6$. $D_{\rm 1, clear}$ is a fixed parameter of the system. Therefore, our auxiliary refractive optics must shorten the telescope's EFFL, translating the focal plane closer to the primary mirror, while minimizing chromatic and geometric aberrations, and interfacing with a CCD dewar. 

We used Zemax OpticStudio to converge on a focal reducer  field-flattener design. We began by modeling the 1.3-m telescope primary and secondary mirrors using the system specifications detailed in Ref. \citenum{Mi96}. We set the distance between the primary and secondary mirror as a variable, bounded by the secondary focus z-travel ($\pm 9.525$ mm). Leveraging previous optical design efforts with similar goals \cite{Ep97}, we began by modeling four spherical lenses in the Zemax Lens Data Editor (LDE), setting their radii (front and back), thicknesses, clear and mechanical semi-diameters as variables to optimize. The distance between the back of the primary mirror and front surface of the first lens was also set as a variable to optimize. 

In addition to defining initial parameter values for all four lenses, we defined a system pressure, temperature, discrete set of wavelengths ($\lambda$), and discrete set of semi-field angles ($u$) for which to optimize the design. We chose a pressure of $0.75$ atm, consistent with the observatory's altitude ($2306$ m), and an operating temperature of $21^{\circ}$C (but see below for a description of our thermal analysis). Our wavelengths of interest were $0.7 \mu\rm{m} \leq \lambda \leq 0.9\mu\rm{m}$ (Section \ref{sec:filter}). Therefore, we optimized our design for three wavelengths within this region, namely $0.7 \mu\rm{m}$, $0.8 \mu\rm{m}$, and $0.9 \mu\rm{m}$. We will use the CCD in frame-transfer mode (Section \ref{sec:ccd}), meaning that our effective imaging area can at most encompass a $0.50 \arcdeg \times 0.25 \arcdeg$ square of sky with a $0.56 \arcdeg$ diagonal. Since we desired optimal image quality across the entire effective imaging area, we chose $0 \arcdeg$, $0.14 \arcdeg$, $0.224 \arcdeg$, and $0.28 \arcdeg$ as the optimization semi-field angles. These correspond to $0\%$ (on-axis), $50\%$, $80\%$ and $100\%$ of the effective radial FOV, respectively. We optimized the design to have a working focal number ($f/\#$) between $5.3$ and $5.6$. The criterion to evaluate optical performance of the system was the root-mean-square (RMS) spot radius at the image plane. 

\begin{figure} [ht]
   \begin{center}
   \begin{tabular}{c}
   \includegraphics[height=6cm]{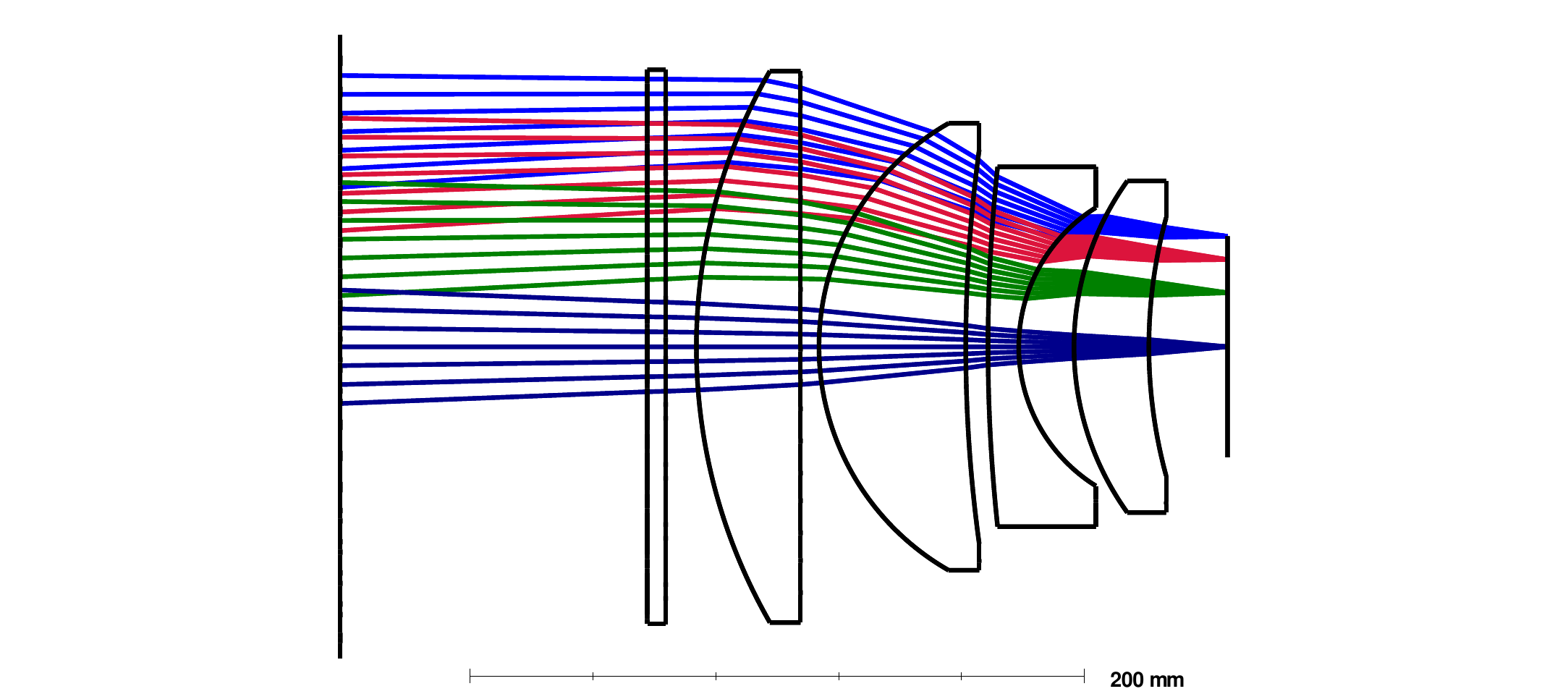} 
   \end{tabular}
   \end{center}
   \caption[example] 
   { \footnotesize Schematic diagram of the \tierras focal reducer field flattener optical design, as seen in cross-section. From left to right:  the primary mirror hole (front surface, vertical solid line), the filter, lens 1 (BSM51Y), lens 2 (CaF2), lens 3 (STIH-6), lens 4 (S-BAH27), and the image plane (vertical solid line). Seven representative rays are shown at each of the four optimization semi-field angles: $0 \arcdeg$ (dark blue), $0.140 \arcdeg$ (green), $0.224 \arcdeg$ (red), and $0.280 \arcdeg$ (royal blue), corresponding to $0\%$ (on-axis), $50\%$, $80\%$ and $100\%$ of the effective radial FOV, respectively. Note how close lens 4 must be to the image plane ($25.61$ mm separation).  \label{fig:optics} }
   \end{figure}

\begin{figure} [ht]
   \begin{center}
   \begin{tabular}{c}
   \includegraphics[height=6cm]{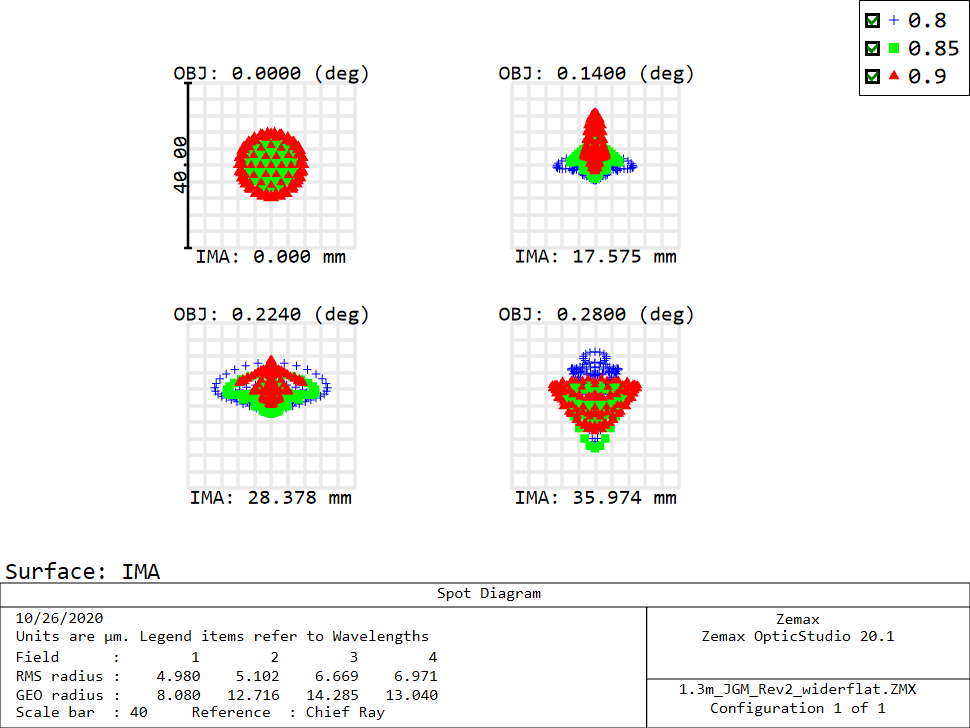}
   \end{tabular}
   \end{center}
   \caption[example] 
   { \footnotesize Spot diagrams of the \tierras focal reducer and field flattener design at four representative semi-field angles: field No. 1 at $0 \arcdeg$ (top left), field No. 2 at $0.140 \arcdeg$ (top right), field No. 3 at $0.224 \arcdeg$ (bottom left), and field No. 4 at $0.280 \arcdeg$ (bottom right). To assess chromatic aberration, each spot is shown at three key wavelengths: $0.8 \mu\rm{m}$ (blue cross), $0.85 \mu\rm{m}$ (green square), and $0.9 \mu\rm{m}$ (red triangle). The RMS and GEO radii for each field (in $\mu$m) are reported in microns in the table underneath the diagrams. The for-fabrication design has an f-number of $f/\# = 5.51$, and a plate scale of $S = 29.25 \arcsec$/mm. It follows that the GEO spot diameters (in arcseconds) are $0.47 \arcsec$ (field 1), $0.74 \arcsec$ (field 2), $0.84 \arcsec$ (field 3), and $0.76 \arcsec$ (field 4). The scale bar for each diagram spans 40 $\mu$m ($1.17 \arcsec$). \label{fig:spots}}
   \end{figure}

We employed the Zemax standard damped least squares algorithm in a series of optimization runs to pinpoint a permutation of lens materials that would yield the tightest and least chromatic spots at each semi-field angle. The combination of BSM51Y (refractive index, $n_D = 1.60311$ and Abbe number, $v_D = 60.65$) for lens 1, CaF2 ($n_D = 1.43384, v_D = 95.23$) for lens 2, STIH-6 ($n_D = 1.80518, v_D = 25.42$) for lens 3, and S-BAH27 ($n_D = 1.70154, v_D = 41.24$) for lens 4 yielded the tightest RMS and geometric (GEO) spot radii, setting our lens materials. None of these glasses are radioactive. Subsequent optimization runs with these materials and other minor adjustments allowed us to converge on a final for-fabrication prescription, with $f/\# = 5.51$, a plate scale of $29.25 \arcsec$/mm, and a FOV side of $0.49 \arcdeg$. Table \ref{tab:presc} describes the for-fabrication optical specifications for each lens, including radii of curvature and mechanical diameters. In Figure \ref{fig:optics}, we show a schematic diagram of the auxiliary optics. Figure \ref{fig:spots} shows spot radii plots at the four key semi-field angles. 

Of particular importance is the tight air space between the back of the fourth lens and the focal plane ($25.61$ mm). The design proved extremely sensitive to this specification, with the spot radii increasing by as much as $50\%$ in response to a 1 mm increase in the separation between these surfaces. Three practical implications arose from this specification: (i) there is no space for a filter between Lens 4 and the image plane, explaining why we put the filter in front of Lens 1 (Figure \ref{fig:optics}); (ii) Lens 4 must act as the CCD dewar window (Section \ref{subsec:optfab}); and (iii) there is no space for a shutter, which partially motivates our use of the CCD in frame-transfer mode (Section \ref{sec:ccd}).

\begin{table}[ht]
\renewcommand{\thetable}{\arabic{table}}
\centering
\caption{\footnotesize Focal reducer and field flattener for-fabrication prescription. Units: mm}  
\label{tab:presc}
\begin{tabular}{ccccc}
Specification  & Lens 1 & Lens 2 & Lens 3 & Lens 4  \\
\hline
\hline
Material & BSM51Y & CaF2 & S-TIH6 & S-BAH27 \\
Radius of Curvature (Front)  & 180.43 & 83.86 & 551.04 & 92.65  \\
Radius of Curvature (Back)   & $\infty$ & 472.98 & 53.46 & 160.00 \\
Center Thickness  & 33.93 & 47.81 & 10.00 & 24.39 \\
Sag (Back) & 0.00 & 4.37 & 25.13 & 5.61 \\
Clear Diameter (Front)  & 173.58 & 139.58 & 111.17 & 85.10 \\
Clear Diameter (Back)  & 168.80 & 122.20 & 84.70 & 78.00 \\
Mechanical Diameter  & 179.58 & 145.57 & 117.17 & 108.01 \\
\hline
\end{tabular}
\end{table}

Prior to proceeding with lens fabrication, we used Zemax to carry out thermal, ghosting, and tolerance analyses of our design. For our thermal analysis, we chose five discrete temperatures between $-10^{\circ}$C and $25^{\circ}$, a range that spans all temperatures the system will encounter in the lab and the site. We used Zemax to model our system at each of these five temperatures while accounting for three main effects: First, the change in the refractive index of each material; second, the change in the radius, thickness and mechanical diameter of each lens due to the material's linear coefficient of thermal expansion (LCTE); and third, the change in the spacing between lenses due to the LCTE of the mounting material (aluminum). The RMS spot radii of our optical design increased by less than $5\%$ for all semi-field angles and all temperatures between $-10^{\circ}$C and $25^{\circ}$, and was readily absorbed by a minor (few micron) refocus of the system at the secondary. 

We carried out a first-pass, double-bounce ghosting analysis in order to assess whether or not rays reflected off of the image plane, and then off any optical surface could ultimately generate significant ghost pupil reflections at the image plane. Ghosts are particularly detrimental for high-precision photometry because they produce halos around the stellar PSF. We identified only one potential ghost, namely a double-bounce reflection that starts at the image plane, bounces off the front surface of lens 3, and hits the image plane with a distance to ghost pupil of $11$ mm. Further optimization of the lens 3 parameters did not ameliorate the ghost reflections. However, we emphasize this ghost is only slightly problematic, and we expect that our high-performance reflection coatings will reduce its potential impacts on our photometric performance. 

We used the Zemax sequential tolerance analysis tool to quantify the effects of potential lens (and filter) manufacturing defects and alignment errors in our optical performance. We distinguish between two types of tolerances: surface tolerances, corresponding to tolerances placed on the lens and filter specifications for the manufacturers, and element tolerances, which set how well-aligned all lenses (and the filter) must be to maintain the desired image quality. The surface tolerances we accounted for included (for each lens): radius, center thickness, mechanical semi-diameter, surface irregularity, gravity displacements, decenter and tilt (between the optical and edge axes of each lens). Our element tolerances were used to quantify the allowable decenter, tilt, and axial misalignment afforded to each optical element during assembly.

The tolerance analysis consisted of two parts: We first carried out a sensitivity analysis to refine the priors for each individual tolerance (in accordance with the lens vendor's capabilities and our in-house alignment experience), and identify the most sensitive parameters. We found decentration, axial displacement, and tilt of lens 2 (CaF2) to be the most stringent tolerances (requiring an error below $\pm 0.035$ mm for each degree of freedom). We then used the refined priors in a Monte Carlo (MC) simulation to statistically predict the impact that all of these tolerances or degrees of freedom (DoF) could simultaneously have on our optical performance. An individual MC run consisted of drawing a value for each DoF from a uniform distribution bounded by the DoF's priors, incorporating it into the optical design, re-focusing the secondary mirror, and computing the resulting RMS spot radii at each semi-field angle. For a total of 101 DoF, 11,000 ($>$ (DoF)$^{2}$) MC runs were required for convergence. The surface and element tolerances outlined in Tables \ref{tab:surftol} and \ref{tab:eltol} guarantee increases in the RMS spot radius (averaged across wavelengths) of $25 \%$ or less for fields 1 and 4, and $12\%$ or less for fields 2 and 3 in $80\%$ of the MC runs. These tolerances were within the manufacturing capabilities of our lens and filter vendors, as well as our in-house assembly capabilities, allowing us to proceed with lens and filter purchasing, and the fabrication of their mechanical supports (Section \ref{subsec:optfab}).

\begin{table}[ht]
\renewcommand{\thetable}{\arabic{table}}
\centering
\caption{\footnotesize A subset of the surface tolerances for the focal reducer and field flattener lenses, provided to the manufacturers. } 
\label{tab:surftol}
\begin{tabular}{ccc}
Tolerance & Value & Units   \\
\hline
\hline
Radius of Curvature (Front and Back) & $\pm 0.02$ & $\%$  \\
Center Thickness & $\pm 0.06$ & mm \\
Sag (Back) & $^{+ 0.25}_{- 00}$ & mm\\
Mechanical Diameter & $\pm 0.05$ & mm \\
Surface Irregularity & 1 & Fringe, Peak-to-Valley (FR PV) \\ 
Surface Figure Error & 0.2 & waves/cm RMS, 1 mm integration length\\
Edge Thickness Difference & 0.025 & mm\\
\hline
\end{tabular}
\end{table}

\begin{table}[ht]
\renewcommand{\thetable}{\arabic{table}}
\centering
\caption{A subset of the element tolerances for our in-house optical alignment and assembly procedure.} 
\label{tab:eltol}
\begin{tabular}{cccccc}
Tolerance  & Filter & Lens 1 & Lens 2 & Lens 3 & Lens 4  \\
\hline
\hline
Axial Displacement (mm) & $\pm$ 0.245 & $\pm$0.080 & $\pm$0.035 & $\pm$0.035 & $\pm$0.080 \\
Decenter (mm) & $\pm$ 0.245 & $\pm$0.080 & $\pm$0.035 & $\pm$0.035 & $\pm$0.080 \\
Tilt (mrad)  & 2.8205 & 0.3491 & 0.1745 & 0.1745 & 0.3491 \\
\hline
\end{tabular}
\end{table}

\subsection{Lens Coatings} \label{subsec:coatings}

\begin{figure} [ht]
\begin{center}
\begin{tabular}{c} 
\includegraphics[height=7cm]{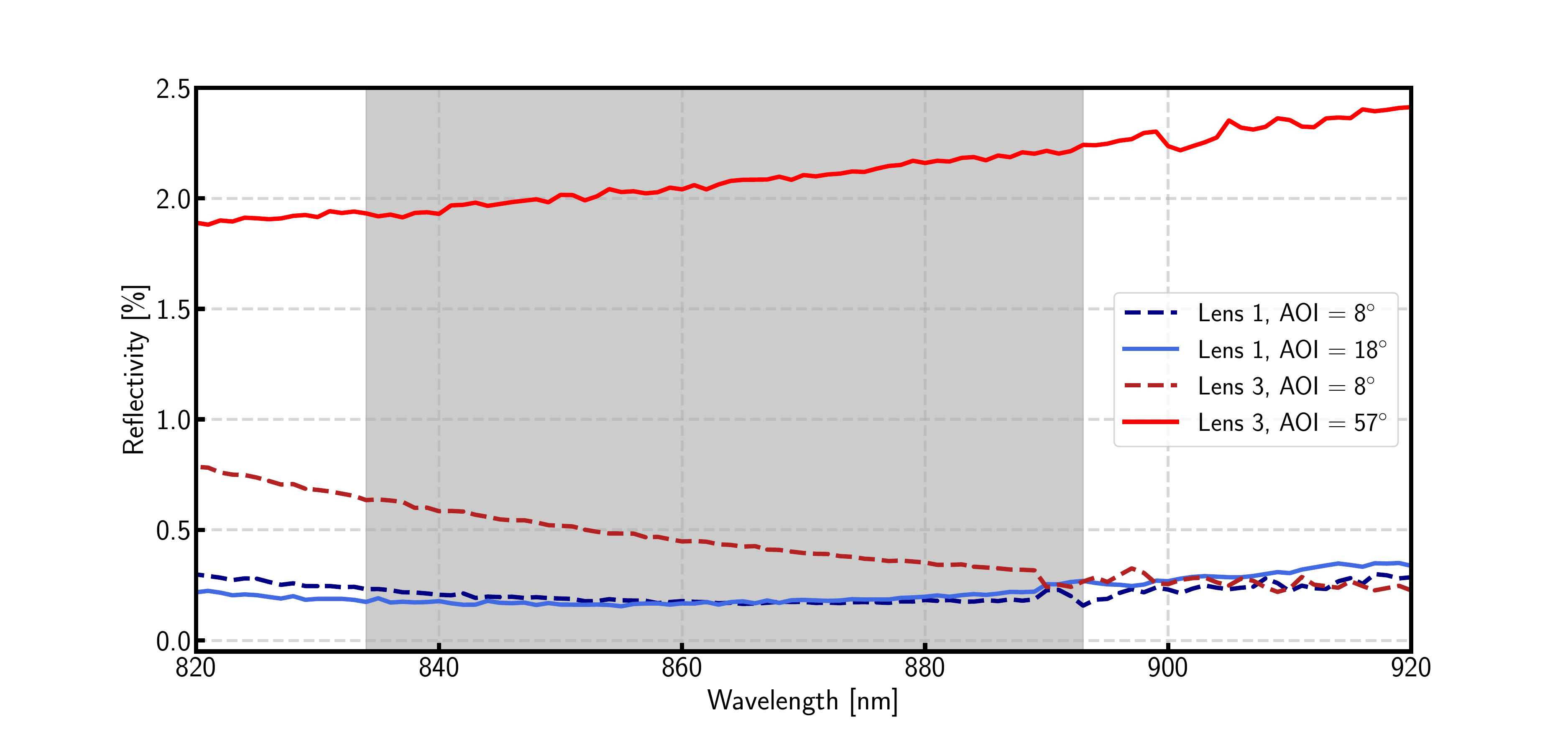}
\end{tabular}
\end{center}
\caption[example] 
{\footnotesize \label{fig:L3coating} Measured percent reflectivity as a function of wavelength for two coated surfaces: the plano surface of lens 1, at AOIs of $8 \arcdeg$ (dark blue, dashed) and $18 \arcdeg$ (royal blue, solid); and the concave surface of lens 3, at AOIs of $8 \arcdeg$ (brick red, dashed) and $57 \arcdeg$ (red, solid). All curves shown correspond to the coating performance at the center of the lens. Due to the pronounced radius of curvature of lens 3's concave surface, the performance of this lens coating at large AOI is the least optimal in our bandpass of interest ($834-893$ nm, shaded gray).}
\end{figure}

The \tierras Observatory will observe in a relatively narrow ($40.2$ nm FWHM) bandpass centered around $863.5$ nm (Section \ref{sec:filter}). Hence, we collaborated with the coating vendor to design and apply high-performance anti-reflective (AR) coatings to the front and back surfaces of our lenses, optimized to transmit maximally in this wavelength range. Table \ref{tab:aois} summarizes the angle of incidence (AOI) range across the entire surface for each of the four lenses, which we calculated using Zemax. The coatings applied to lenses 1,2, 4 and the convex surface of lens 3 have a maximum reflectivity, $R_{\rm max}= 0.75\%$ for all AOIs ($0-38^{\circ}$). The concave surface of lens 3 was the most challenging to coat due to its pronounced radius of curvature. For this surface, $R_{\rm max} = 0.75\%$ for AOI between $0-45^{\circ}$, while $R_{\rm max} = 2.5\%$ for AOI between $45-57^{\circ}$. The reflectivity uniformity, $\Delta R$, for all coated surfaces varies by $<0.25 \%$ from center to edge, at any given wavelength and AOI. All of the above specifications apply to the $834-893$ nm wavelength range. However, our AR coating is also moderately transmissive in the visible range, with an average reflectivity, $R_{\rm avg}$, below $5.0\%$ in the $450-650$ nm wavelength range for all surfaces. This design choice allows for comfortable inspection, testing, alignment, and assembly using visible light sources in the lab. Figure \ref{fig:L3coating} exemplifies the wavelength-dependent performance of our lens coatings for representative AOIs in our bandpass of interest.

\begin{table}[ht]
\renewcommand{\thetable}{\arabic{table}}
\centering
\caption{\footnotesize Range of angles of incidence for each lens (front and back) surface in the \tierras focal reducer and field flattener.} 
\label{tab:aois}
\begin{tabular}{ccc}
Surface & Clear Semi-Diameter (mm) & AOI range ($^{\circ}$) \\
\hline
\hline
Lens 1 (Front) & 86.8 & $0 - 27$\\
Lens 1 (Back) & 84.4 & $0 - 18$ \\
Lens 2 (Front) & 69.8 & $0 - 38$\\
Lens 2 (Back) & 61.1 & $0 - 35$\\
Lens 3 (Front) & 55.6 & $0 - 35$\\
Lens 3 (Back) & 42.4 & $0 - 57$\\
Lens 4 (Front) & 42.6 & $0 - 36$\\
Lens 4 (Back) & 39.6 & $0 - 16$\\
\hline
\end{tabular}
\end{table}

\subsection{Current Status: Opto-Mechanical Design, Fabrication and Assembly} \label{subsec:optfab}

We provided the lens vendor (Optimax Systems Inc) with lens blanks from OHARA (BSM51Y, STIH-6, S-BAH27) and HELLMA (CaF2) in 2019 June. OHARA provided melt data for the three lens blanks, which we used to refine our material models within Zemax, re-balance the design, and provide Optimax with the final for-fabrication prescription (Table \ref{tab:presc}). Our four lenses were polished, coated, and delivered in 2020 January. The lenses met our required surface tolerances (Table \ref{tab:surftol}). The \tierras focal reducer and field flattener as-built specifications, summarized in Table \ref{tab:ABpresc}, ultimately yield a $f/\# = 5.6$ system with a FOV side of $0.48 \arcdeg$ and a plate scale of $28.82 \arcsec$/mm. 

\begin{table}[ht]
\renewcommand{\thetable}{\arabic{table}}
\centering
\caption{\footnotesize Focal reducer and field flattener as-built lens specifications. Units: mm}  
\label{tab:ABpresc}
\begin{tabular}{ccccc}
Specification  & Lens 1 & Lens 2 & Lens 3 & Lens 4  \\
\hline
\hline
Radius of Curvature (Front)  & 180.419 & 83.870 & 551.004 & 92.663  \\
Radius of Curvature (Back)   & $\infty$ & 472.987 & 53.456 & 159.993 \\
Center Thickness  & 33.966 & 47.803 & 10.034 & 24.412 \\
Sag (Back) & 0.000 & 4.465 & 25.180 & 5.708 \\
Mechanical Diameter  & 179.595 & 145.607 & 117.212 & 108.046 \\
\hline
\end{tabular}
\end{table}

\begin{figure} [ht]
\begin{center}
\begin{tabular}{c} 
\includegraphics[height=8cm]{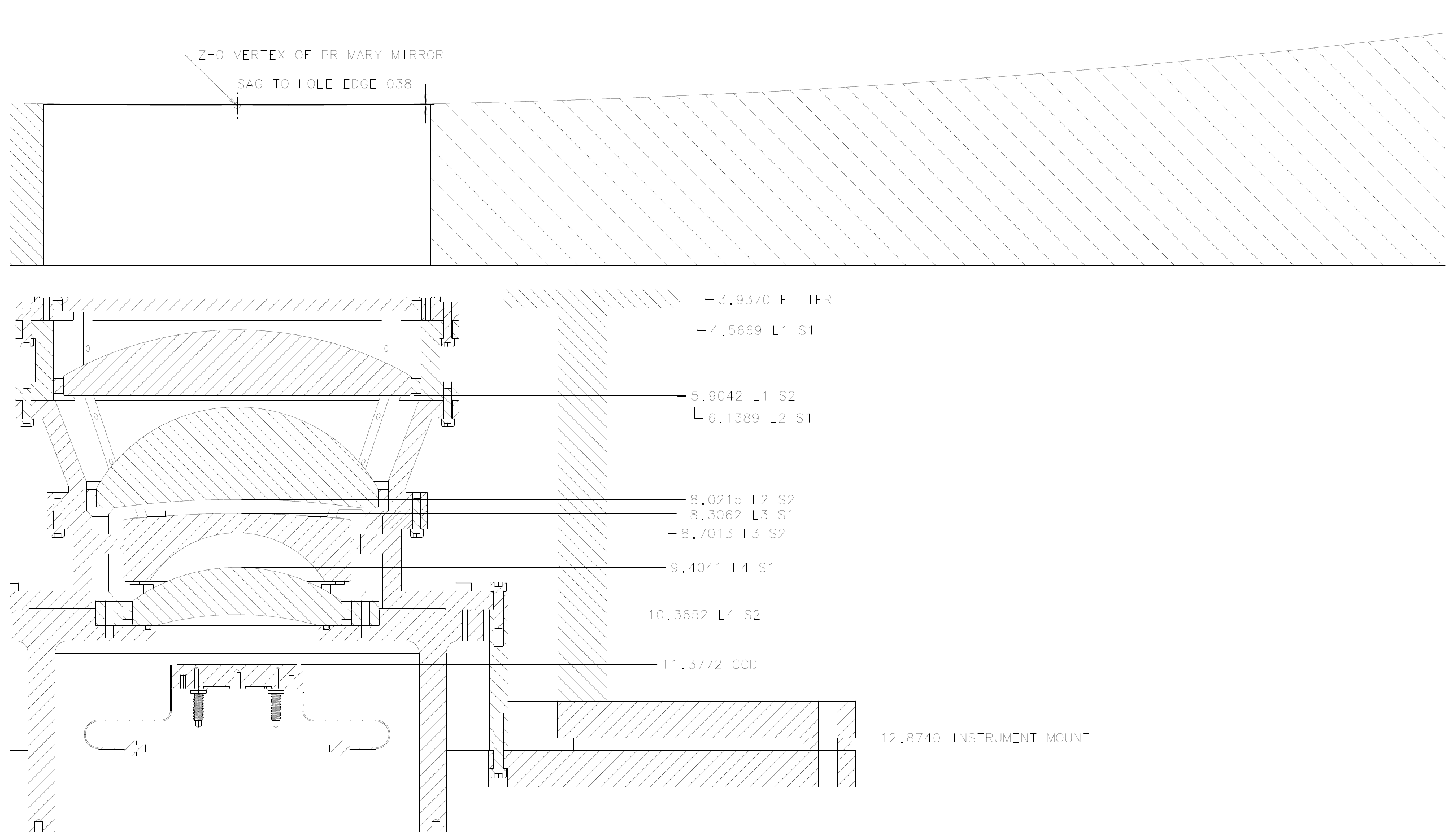}
\end{tabular}
\end{center}
\caption[example] 
{\label{fig:assy} \footnotesize Opto-mechanical assembly of the \tierras focal reducer, field flattener, and CCD dewar. The distance of critical surfaces to the primary mirror vertex is shown for reference.}
\end{figure}

\begin{figure} [ht]
\begin{center}
\begin{tabular}{c} 
\includegraphics[height=7cm]{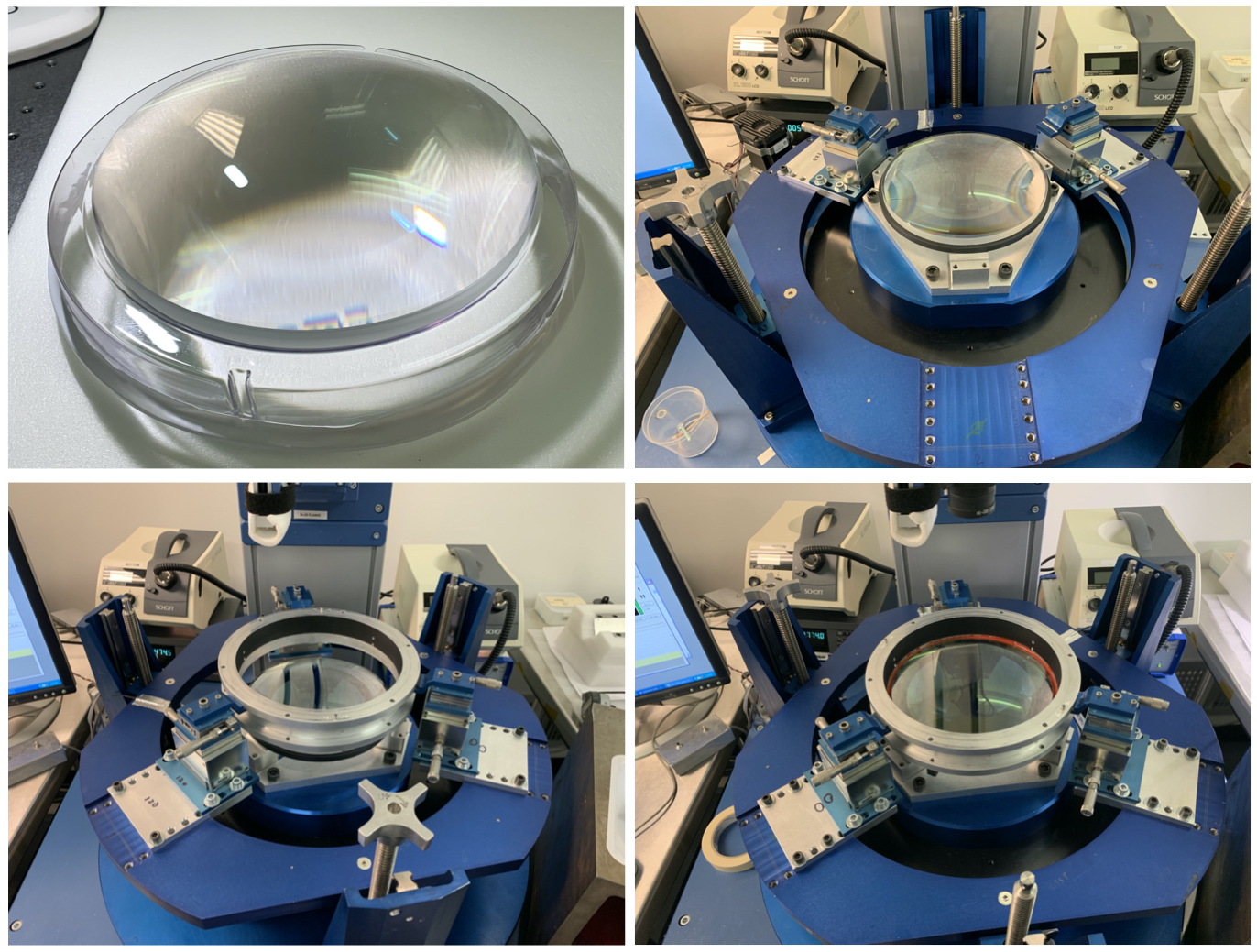}
\end{tabular}
\end{center}
\caption[example] 
{\footnotesize \label{fig:bonding} The lens alignment and bonding process for lens 1 (BSM51Y) inside our clean room. { \em Top left:} polished and coated lens sitting above its delivery casing bottom. After removing the lens from its casing, we cleaned and primed its edge with MOMENTIVE SS4155 blue primer. {\em Top right:} lens sitting on top of the TRIOPTICS Opticentric high-precision lens centration and alignment machine's tilt/translation table (blue and black circular surfaces below lens). Note the lens sits inside a black delrin bond ring and interfaces to the Opticentric table via custom aluminum fixturing. The delrin ring is meant to prevent liquid RTV spillage during bonding. {\em Bottom left:} cleaned and primed bezel sits above the lens and is independently supported by three fine adjustment centration and tilt blocks. The blocks are isolated from the Opticentric tilt/translation table to enable independent bezel and lens alignment. In subsequent steps, the bezel is carefully lowered and positioned optimally with respect to the lens vertex in order to fulfill our optical prescription and axial displacement/centration/tilt tolerances. {\em Bottom right:} After the lens and bezel are optimally positioned with respect to each other, we deposit MOMENTIVE RTV560 (red liquid rubber) between the lens edge and the bezel to bond them together. }
\end{figure}

We finalized the design, drafting, and building of all opto-mechanical support components for the lenses in 2020 February. Figure \ref{fig:assy} lays out the entire opto-mechanical assembly below the 1.3-m primary mirror, including all bezels and fixturing as well as the filter and the CCD inside its custom dewar. Each lens is attached to an aluminum bezel with a ring of RTV560, a two-part silicone rubber that retains elastometric properties in a wide temperature range ($-115^{\circ}$C to $260^{\circ}$C). With a LCTE of $20 \times 10^{-5}$ cm/cm/$^{\circ}$C, RTV560 complements the material properties of aluminum (LCTE$=24 \times 10^{-6}$ cm/cm/$^{\circ}$C) to guarantee that each lens remains centered and aligned in its respective bezel within our allowable tolerance budget (Table \ref{tab:eltol}) as the dome temperature changes throughout a night of observation. Note that lens 4 acts as the CCD dewar window. We are currently bonding lenses inside their respective bezels in the lab, as shown in Figure \ref{fig:bonding}.

\section{The \tierras Filter} \label{sec:filter}

\subsection{Design and Fabrication} \label{subsec:filtdsgn}

Previous efforts with state-of-the-art observatories have shown that ground-based photometry of red dwarfs is limited by changes in the precipitable water vapor at the observing site, since water features in the photometric bandpass result in a color dependent offset when comparing the relative brightness of the red target to the blue field stars \cite{Be12}. Motivated by this context, we set out to determine whether there existed a combination of filter parameters in the $700 - 900$nm wavelength range that could minimize the contribution of PWV changes towards the total photometric error budget (Eq. \ref{eq:toterr}), while maximizing the amount of stellar photons received from our future mid-to-late M dwarf targets. The filter candidates' parameters consist of a center wavelength $\lambda_C$, and maximum transmission bandpass $\Delta \lambda$.

We set up a simulation to quantify the PWV error (Eq. \ref{eq:pwverr}) for an ensemble of box-shaped filter candidates with $845$ nm$ < \lambda_C < 885$ nm and $30$ nm $< \Delta \lambda < 70$ nm. We used observed and calibrated HST Calspec spectra\cite{Bo14} of Gliese 555 (M3.5V) and vb8 (M7Ve) to explore the effect of spectral type on PWV error. In both cases, we used 18 Sco (spectral type G2Va) as our comparison star. To determine the flux $F$ (in ergs s$^{-1}$ cm$^{-2}$) of our target and comparison stars through the $n^{th}$ filter being evaluated at a given PWV, we integrated the product of the overall system transmission $T(\lambda, \rm PWV)$ and the stellar spectrum $f(\lambda)$ as observed from Earth (in ergs s$^{-1}$ cm$^{-2}$ \AA$^{-1}$). $T(\lambda, \rm PWV)$ is given by the product of the empirical quantum efficiency curve of our CCD (Section \ref{sec:ccd}), the transmission curve of the $n^{th}$ filter being evaluated, the optical transmission fraction due to the telescope mirrors and coated auxiliary optics (Section \ref{sec:optics}), and the telluric spectrum \cite{Be14} scaled to the desired mm of column water vapor. 

\begin{figure} [ht]
\begin{center}
\begin{tabular}{c} 
\includegraphics[height=5.3cm]{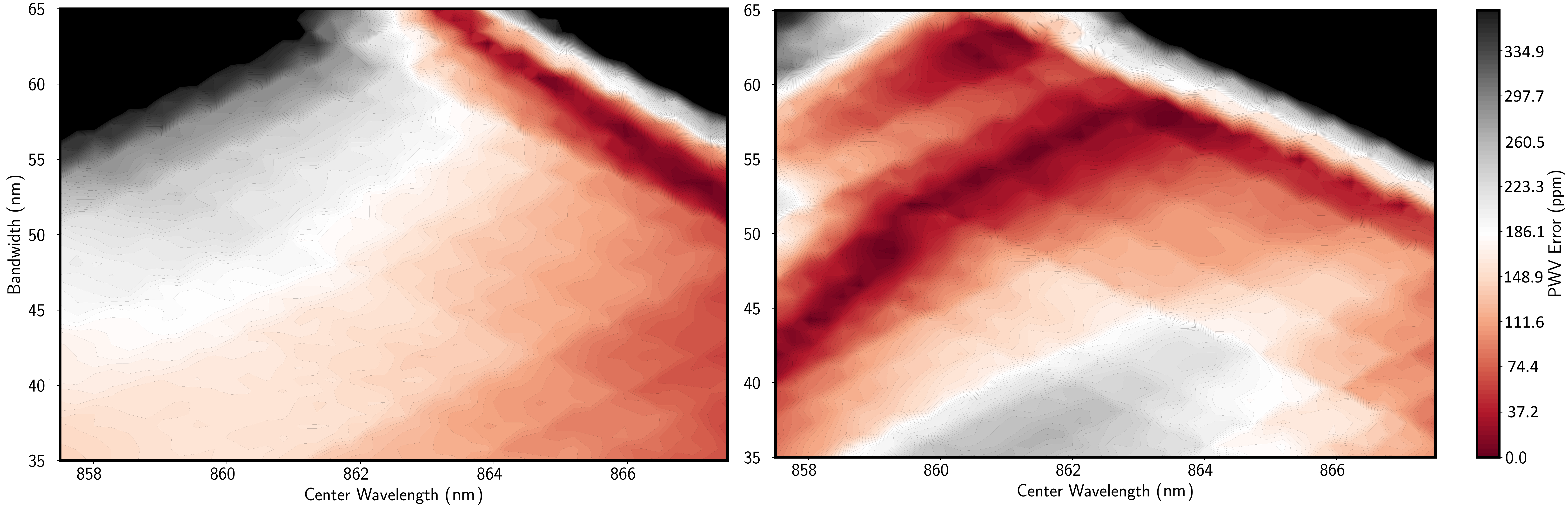}
\end{tabular}
\end{center}
\caption[example] 
{\footnotesize \label{fig:pwv3d} Expected precipitable water vapor error (in ppm) through the \tierras system as a function of filter bandwidth $\Delta \lambda$ (nm) and center wavelength $\lambda_C$ (nm). {\em Left:} Gliese 555 (M3.5V) as the target star. {\em Right:} vb8 (M7Ve) as the target star. Although the simulation spanned a wider range of bandwidths and center wavelengths, the plot zooms into the region of greatest interest. Filter candidates with $\Delta \lambda < 45$ nm are disfavored as they yield very high photon noise. For reference, the MEarth filter has $\lambda_C = 875$ nm and $\Delta \lambda = 250$ nm, placing it within a region of parameter space, which is not shown in the figure, where the PWV error is much greater than $364$ ppm.
\vspace{5mm}}
\end{figure}

Figure \ref{fig:pwv3d} quantifies the expected PWV error for the mid M (Gliese 555) and late M-dwarf (vb8), zooming into the zone of greatest interest ( $857.5$ nm$ < \lambda_C < 867.5$ nm and $35$ nm $< \Delta \lambda < 65$ nm). Whereas for the later M-dwarf a wider range of $\lambda_C$ (858 to 867 nm) and $\Delta \lambda$ (40 to 65 nm) should allow one to reach a very low PWV noise floor, for the earlier M-dwarf one must constrain the filter to have $863$ nm$ < \lambda_C < 885$ nm and $50$ nm $< \Delta \lambda < 70$ nm. Figure \ref{fig:pwv3d} hints at an additional challenge to blindly following the filter constraints placed by the earlier M-dwarf: a manufacturing error in $\lambda_C$ of 1-2 nm, or in $\Delta \lambda$ of 4 nm, would shift the filter performance to a region in parameter space where the expected PWV error for the late M-dwarf would increase well above 364 ppm. Thus, it became imperative to carry out a second simulation to model more realistic filter candidates that account for manufacturing limitations. In this case, each filter candidate was parameterized by: a center wavelength and error, $\lambda_C \pm \delta \lambda_C$; a slope width, $\Delta s$, for the red and blue edges of the filter; a transmission bandpass,$\Delta \lambda$, measured from the $50\%$ transmission wavelengths in the red and blue slopes of the filter; and two wavelengths in the blue ($\lambda_B$) and red limits ($\lambda_R$) where a certain attenuation level ($T_{\rm up}$) had to be reached. Table \ref{tab:filtrange} summarizes the range of values explored for each of these parameters. 

\begin{table}[ht]
\renewcommand{\thetable}{\arabic{table}}
\centering
\caption{\footnotesize Allotted value range for each filter parameter.} 
\label{tab:filtrange}
\begin{tabular}{cccc}
Parameter & Description &Value range & Units \\
\hline
\hline
$\lambda_C$ & Center wavelength & $830-910$ & nm \\
$\delta \lambda_C$ & Center wavelength error & $\pm 10$ & nm \\
$\Delta s$ & Slope width & $3-10$ & nm \\
$\Delta \lambda$ & FWHM Bandwidth & $10-70$ & nm \\
$\lambda_B$ & Blue wavelength where $T \leq T_{\rm up}$ & $820-840$ & nm\\ 
$\lambda_R$ & Red wavelength where $T \leq T_{\rm up}$ & $880-920$ & nm\\
$T_{\rm up}$ & Attenuation upper-bound & $0.1 - 0.5$ & $\%$\\
\hline
\end{tabular}
\end{table}

\begin{figure} [ht]
\begin{center}
\begin{tabular}{c} 
\includegraphics[height=7cm]{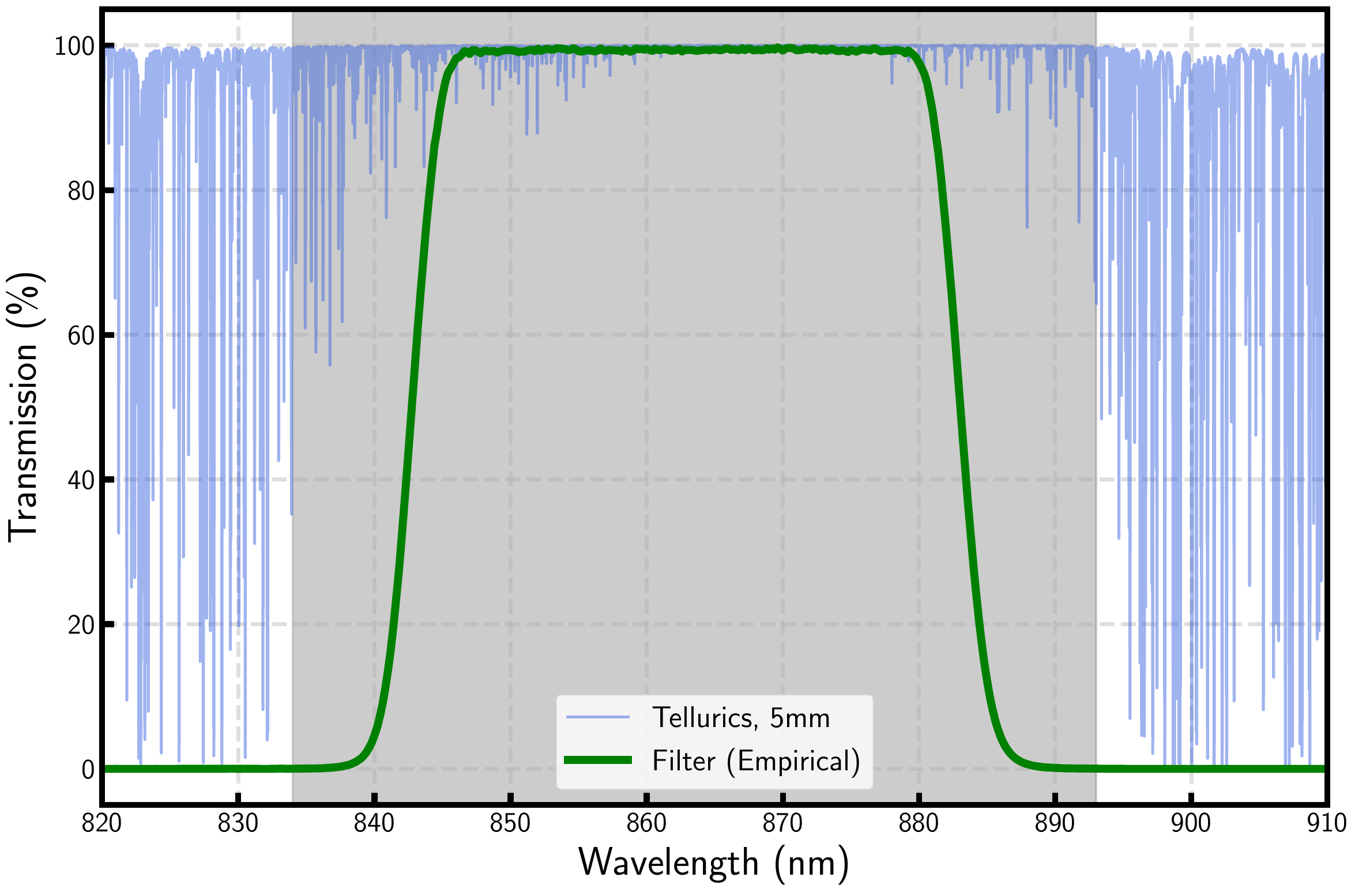}
\end{tabular}
\end{center}
\caption[example] 
{\footnotesize \label{fig:filtT} Filter empirical transmission curve, shown in green. The Earth's telluric spectrum is shown in blue (scaled to have a 5mm column water vapor). Attenuation levels below $0.1\%$ are reached outside the region shaded gray. Specifically, $T < 0.1 \%$ between $300-834$ nm and $893-1100$ nm.}
\end{figure}

We found that the single most important filter specification to minimize the PWV error in our observations consisted of reaching attenuation levels $T \leq T_{\rm up}= 0.1\%$ for wavelengths $\leq 834$ nm and $\geq 893$ nm (Figure \ref{fig:filtT}), while maximizing transmission interior to this wavelength range. Furthermore, this attenuation specification must to be met for the entire filter area and for all of the filter AOIs ($0-4^{\circ}$). Maximizing transmission entails maximizing the width of the region interior to $834-893$ nm that reaches $T>95\%$ and peak-to-valley variation within $\pm 5\%$, which is enabled by having the steepest possible slopes on the red and blue limits of the filter in wavelength space. We collaborated with the filter vendor, Asahi Spectra Co., Ltd., to determine the steepest slopes they could achieve while guaranteeing the aforementioned attenuation requirement. We did not require the filter slopes to be symmetrical, and in fact favored a steeper slope on the red side of the filter. Figure \ref{fig:filtT} shows the empirical transmission curve for our filter, which Asahi provided along with the filter in 2020 June. Table \ref{tab:filtspecs} summarizes all other relevant filter specifications.

\begin{table}[ht]
\footnotesize
\renewcommand{\thetable}{\arabic{table}}
\centering
\caption{\footnotesize Select filter specifications given to Asahi. They were all met or exceeded.}
\label{tab:filtspecs}
\begin{tabular}{ccc}
Specification Category & Specification & Description \\
\hline
\hline
Operating Conditions & Focal ratio & $f/13.5$  \\
& Angles of incidence & $0-4^{\circ}$ \\ 
& Beam diameter at filter & 3.6 cm \\
& Operating temperature range & -10 to 25$^{\circ}$C \\
& Operating humidity & 0 to 95\% \\
& Orientation to gravity & All orientations are possible \\
& Operating Pressure & 540-580 torr \\
\hline 
Optical Specifications & Center Wavelength ($\lambda_C$) & $863.5 \pm 3 $ nm at several filter locations \\
& FWHM & $40 \pm 3$ nm \\
& BW90 ($.9\times$ width of peak transmission) & $> 0.85$FMHW \\
& BW10 ($.1\times$ width of peak transmission) & $< 1.15$FMHW \\
& Out-of-band Transmission & $<0.1\%$ for 300-834 nm, 893-1100 nm \\
& Transmission Uniformity & $\pm 2\%$ of avg., entire clear aperture\\ 
\hline
Physical Specifications & Material & High-quality Fused Silica\\
& Thickness & $6.00^{+.20}_{-.00}$ mm\\
& Mechanical Diameter & $180.41 \pm .30$ mm \\
& Min Clear Aperture & $174.41$ mm\\
& Surface Parallelism & $\leq 1 \arcmin$\\
& Coatings & Both surfaces AR-coated. \\
& Transmitted wavefront & $<\lambda_C/4$ in any 36 mm sub-aperture \\
\hline 
\end{tabular}
\end{table}

\subsection{Mechanical Assembly}

\begin{figure} [ht]
\begin{center}
\begin{tabular}{c} 
\includegraphics[height=6cm]{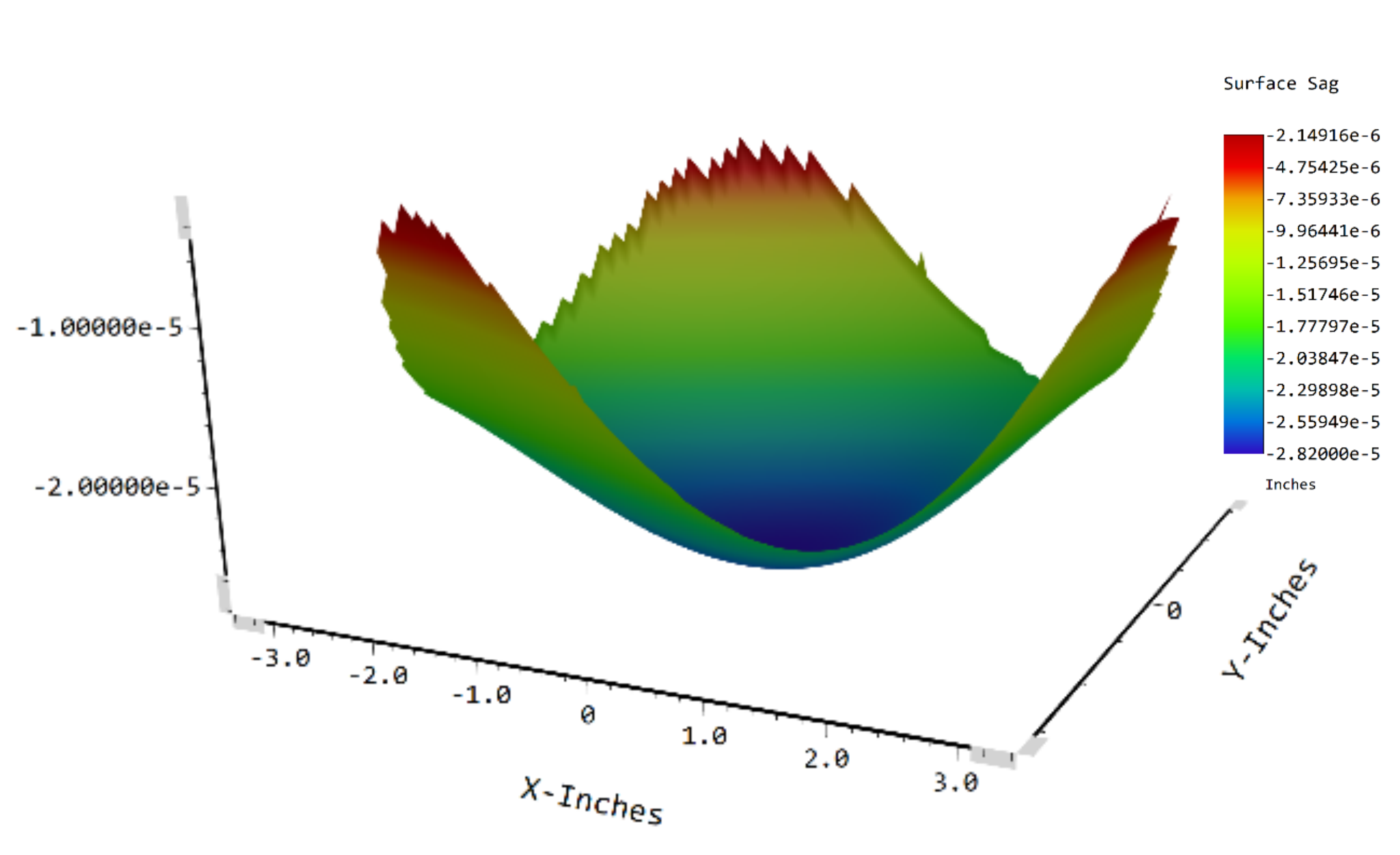}
\includegraphics[height=5cm]{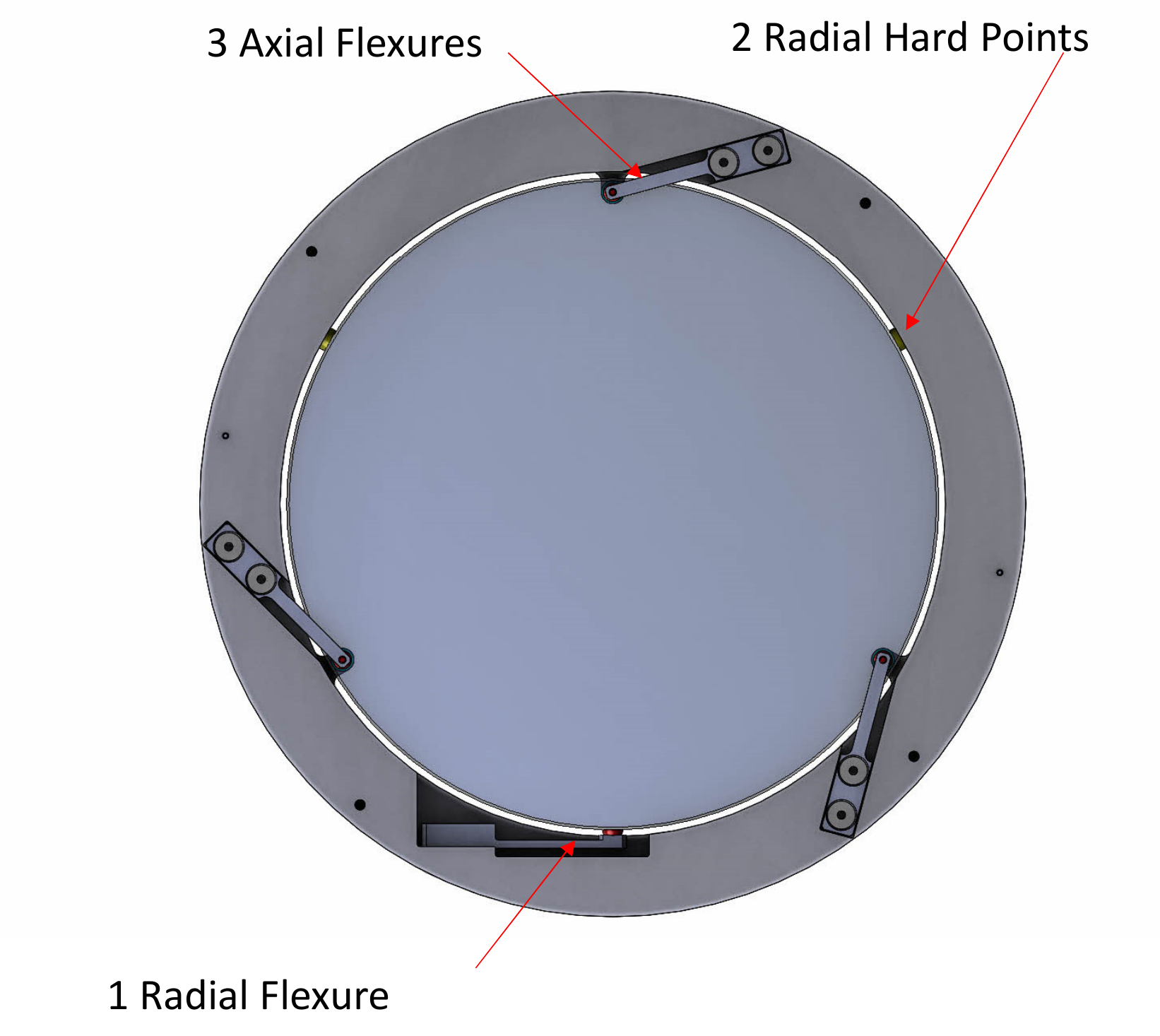}
\end{tabular}
\end{center}
\caption[example] 
{\label{fig:filtflex} \footnotesize {\em Left:} 3D surface sag model quantifying the expected spatial filter self-weight deformation assuming it is held radially at three points (red). {\em Right:} Bird's eye view of the flexured filter mount. The three axial flexures, radial flexure, and two radial hard points are labelled for clarity.}
\end{figure}

In lieu of attaching the filter to its aluminum bezel with RTV560, we opted for a three-point flexured mount (Figure \ref{fig:filtflex}). This design choice was motivated by a concern that the silicone rubber might damage the filter after prolonged exposure. We used Zemax to assess the effect of self-weight deformation (caused by holding the filter axially from three points) on optical performance when pointing the telescope towards the zenith. We did not consider other orientations since we expect the largest amount of filter deformation to occur when the telescope is pointing towards the zenith. We modelled the sag, $z(\rho, \phi)$, of the front and back surfaces of the filter by means of a Zernike polynomial of the form $$z(\rho, \phi) = A_1 + A_4(2 \rho^2 - 1) + A_{10}\rho^3 \cos 3 \phi $$ where $\rho$ is the normalized radial ray coordinate, $\phi$ is the azimuthal ray coordinate, and $A1$, $A_4$, $A_{10}$ are the Zernike coefficients representing piston (mean value of the wavefront or phase profile across the pupil), defocus, and trefoil. We used a set of boundary conditions derived from a separate gravity displacement analysis to derive $A_1 = -1.9 \times 10^{-5}$, $A_4 = 9.2 \times 10^{-6}$, and $A_{10} = 9.8 \times 10^{-6}$. After deforming the filter, we inspected the spot image quality for the four key semi-field angles and found no appreciable increase in their RMS or GEO radii. Thus, we proceeded with designing, drafting, and fabricating the flexured bezel mount and accompanying fixturing. We are in the process of assembling the flexured bezel mount and filter in the lab.

\section{Deep-Depletion CCD and Custom Dewar} \label{sec:ccd}

\begin{figure} [ht]
\begin{center}
\begin{tabular}{c} 
\includegraphics[height=7cm]{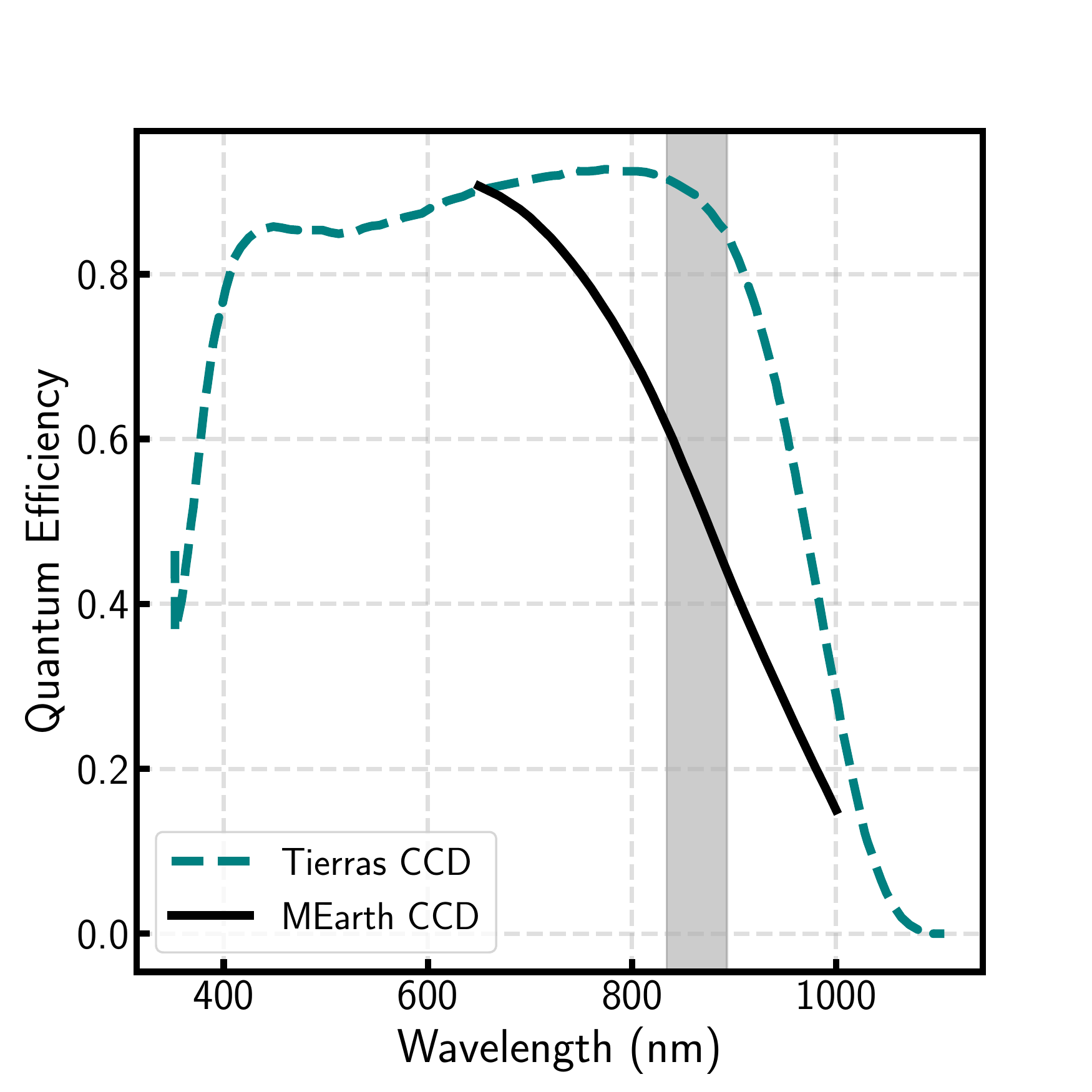}

\includegraphics[height=7cm]{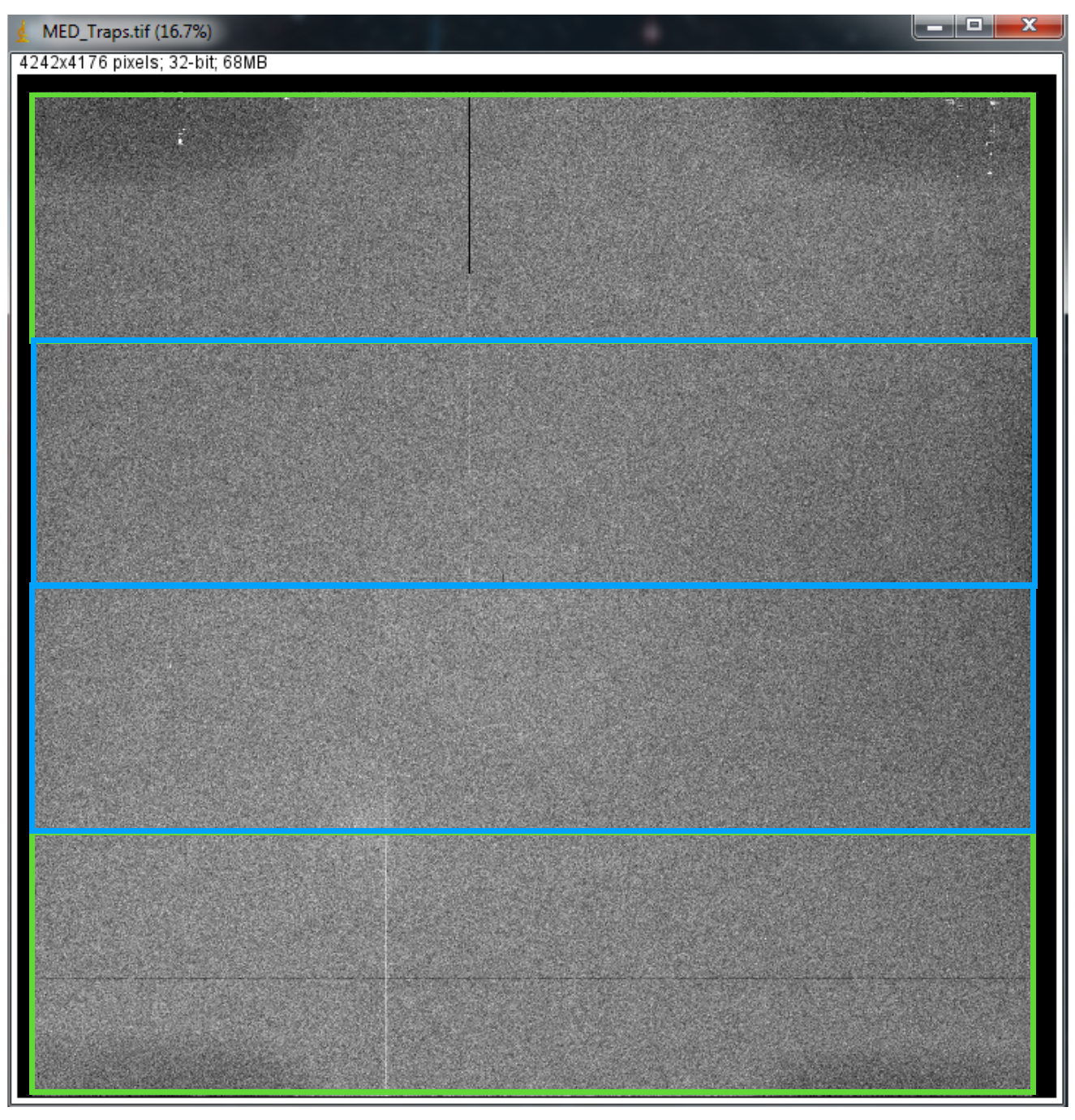}
\end{tabular}
\end{center}
\caption[example] 
{\footnotesize \label{fig:ccdqe} {\em Left: }Empirical Quantum Efficiency (QE) curve of the \tierras NIR-optimized deep-depletion CCD (turquoise). The MEarth CCD QE curve (black) is shown for comparison. The \tierras filter transmission is represented by the shaded (grey) region. {\em Right:} Snapshot of the most salient cosmetic defects of our CCD chip, overlayed with a graphical representation of the effective imaging (blue) and charge transfer (green) pixel areas of the CCD. The readout amplifiers we will use (not shown) are located in the top left and bottom right corners of the chip. }
\end{figure}

\tierras will employ a $4096(H) \times 4112(V)$ Teledyne e2v CCD231-84-*-F64 bulk silicon deep-depletion chip. The device has a pixel size of $15 \, \mu$m, and a full-frame $61.4$ mm $\times 61.7$ mm imaging area. We will use the device in frame-transfer mode since we have no room for a shutter in the optical train, and our application will require a number of shutter closures far in excess of the mean number between failures for commercially available shutters. Thus, our effective imaging area consists of $4096(H) \times 2048(V)$ pixels in the middle of the chip, while the upper and lower $4096(H) \times 1032(V)$ frame transfer areas are covered with a silicon mask. Pixel readout will occur through two amplifiers on opposite sides of the imaging area, with a readout rate of $1.4$ MHz and a read noise of 15-16 e- RMS, quantified in the lab. Two additional binning modes at $1.0$ MHz (12-13 e- RMS) and $500$ kHz (7.5 e- RMS) are also available. Figure \ref{fig:ccdqe} shows the empirical quantum efficiency (QE) curve of our device ( $>85\%$ in the \tierras bandpass), and illustrates the CCD architecture, charge traps, and other defects. The pixel full well signal is 300k e-, with a parallel shift time of $100 \, \mu$sec/row to accommodate the large full well. The CCD sits inside a custom-made camera dewar with the S-BAH27 lens acting as the dewar window. The CCD will operate at $-110^{\circ}$C, cooled using an Edwards Polycold Compact Cooler (PCC). Our PCC uses a PT-30 refrigerant blend (a proprietary combination of argon, nitrogen, methane and propane) as the coolant. This cooling system is closed loop and does not require refills, as do liquid nitrogen dewars, facilitating our automation and remote operation efforts. The custom camera dewar, cooling system, silicon mask, electronics, and control software were designed and built by Spectral Instruments, and delivered in 2020 July. 

CCD testing in the lab is currently underway and will quantify the bias offset and gain stability over time, bright and dark defects (pixels and columns), dark current (e-/pixel/second), Photo-Response Non-Uniformity (PRNU), charge transfer efficiency (via edge pixel extended response method), output amplifier linearity (\% vs signal level), image persistence, relative QE, charge traps, fixed pattern noise (in bias offset), and fixed response patterns (‘tree rings’). 

\section{Telescope Refurbishment} \label{sec:refurb}

The \tierras Observatory must be fully robotic in order to maximize our time on-sky pursuing targets. A first step towards roboticizing the telescope consisted of refurbishing its control system and control system hardware. Telltale signs of the telescope's deterioration included: (i) severe abrasion of the Heidenahin LIDA 105 incremental RA and Dec encoder tapes and read heads, which resulted in loss of telescope position and improper tracking; (ii) a west-side declination bearing that experienced lateral motion and caused non-repeatable multi-arcsecond jumps as the telescope moved in RA; and (iii) an outdated telescope control system (Comsoft PC-TCS) hosted in mostly obsolete ISA slot equipped computers running DOS/Windows 9x. 

We hired DFM Engineering Inc. to carry out the telescope refurbishment during January 2020, which included the following tasks : 

\textit{ Replacement of RA and Dec Encoder System --} The original Heidenhain incremental RA and Dec encoder system was replaced with Renishaw (BiSS interface) 26 bit on-axis absolute RA and Dec encoder read heads and tapes. This required that the telescope's original RA and Dec secondary drive speed reducer assemblies be adapted with new DC servomotor/motor encoder units, as well as custom brackets for mounting the Renishaw read heads. In preparation for encoder tape and read head installation, we cleaned, polished, and rust-treated the telescope's RA and Dec drive disks with WD-40. The new absolute encoders will enable stable pointing and precise tracking. While replacing the read heads and tapes, we also increased the telescope's RA range of motion from $\pm 4.35$ hours to $\pm 5.0$ hours by modifying the location of the limit switches.

\textit{ TCS Upgrade --} DFM replaced the system's original Comsoft PC-TCS with their new proprietary TCSGalil control system, named after the Galil 4183 digital motor controller (DMC) that enables the use of modern (non-ISA slot equipped) computers for telescope control. The TCSGalil computer consists of an Intel Core i7 CPU running Windos 10 OS, and housed inside an iStar 4U industrial rackmount with a wired serial com port, parallel port, network interface, 8 GB ram, and 500 GB hard drive. The DMC and TCS interface via a dedicated TCP/IP port. The TCSGalil software drivers for telescope, focus, and dome control are ASCOM-compliant.

\textit{ Secondary Focus and Tip/Tilt --} The 2MASS survey carried out the first large-scale implementation of freeze-frame scanning to efficiently observe $6 \arcdeg$ declination strips of sky every 7.8 seconds \cite{Sk06}. This technique consists of tilting (colloquially known as ``chopping") the secondary mirror as the telescope scans in declination in order to freeze the focal plane. In turn, the original 1.3-m telescope secondary mirror focus mechanism consisted of three TCS-controllable motors: one focus motor (Kollmorgen P22NRXC-LNF-NS-00 Stepper Motor) and two tilt motors for chopping (BEI Kimco LA13-12-000A linear voice coil actuators). There was no tip nor centration capability. DFM retrofitted the existing Kollmorgen focus stepper motor with a Renishaw 26 bit linear absolute focus encoder to allow for $\pm 9.525$ mm in secondary mirror (z-direction) travel, with position resolution at the level of a few micrometers. Such a fine focus resolution will enable us to obtain the best possible image quality from the telescope and auxiliary optics. The chopping secondary mechanism was replaced with a secondary tip/tilt/translate mechanical collimation assembly between the focus ram and the secondary mirror cell. Secondary tip/tilt, and centration adjustments are carried out by manually turning and locking screws on the assembly. Since we intend the observatory to be robotic, locking the secondary in place will guarantee that we sustain optimal image quality over long periods of time. 

\textit{ Dome Control Refurbishment --} The 1.3-m telescope is housed inside a 22 ft 6 in Model M Ash Dome with lower and upper 6ft shutters, installed circa 1995. DFM replaced the original dome Azimuth motor with AC servo-motors, interfaced to a new variable frequency box to allow for remote and programmable dome slew speed control. A new DFM Ash Dome Azimuth encoder drive and accompanying actuator ramp allow the Galil DMC (and therefore TCSGalil) to command the dome's azimuthal motion and find dome home position. Lastly, DFM added radio-commanded relays to control both dome shutters via TCSGalil.  

Following commissioning of the \tierras auxiliary optics and camera in early 2021, we plan to automate the observatory for remote scheduling and operations using modified software from the MEarth Project.

\section{Expected Performance and Project Status}\label{sec:perf}

To predict our expected system performance and determine whether all the aforementioned design choices will allow us to meet our photometric precision requirements, we estimated our photometric error (Eq. \ref{eq:toterr}) as a function of integration time when observing either an M3.5V (Gliese 555) or an M7Ve(vb8) target star, using a single G2V (18 Sco) comparison star to quantify PWV error. As in Section \ref{sec:filter}, we use HST Calspec spectra scaled to 7.5 and 15 pc for both stars. In our calculation, we also account for the expected throughput of our system ($58\%$) given the measured reflectivity of the primary and secondary mirror, the reflectivity of the coated optics \ref{subsec:coatings}), the transmission of our coated filter (Section \ref{subsec:filtdsgn}), the empirical transmission curve of the \tierras filter (Figure \ref{fig:filtT}), and the empirical QE curve of our deep-depletion CCD (Figure \ref{fig:ccdqe}). We assume an airmass of $\chi = 1.5$ to estimate scintillation, but emphasize that this should only be taken as an approximation of this notoriously hard to model error contribution. 

\begin{figure}[ht]
\begin{center}
\begin{tabular}{c} 
\includegraphics[height=6.83cm]{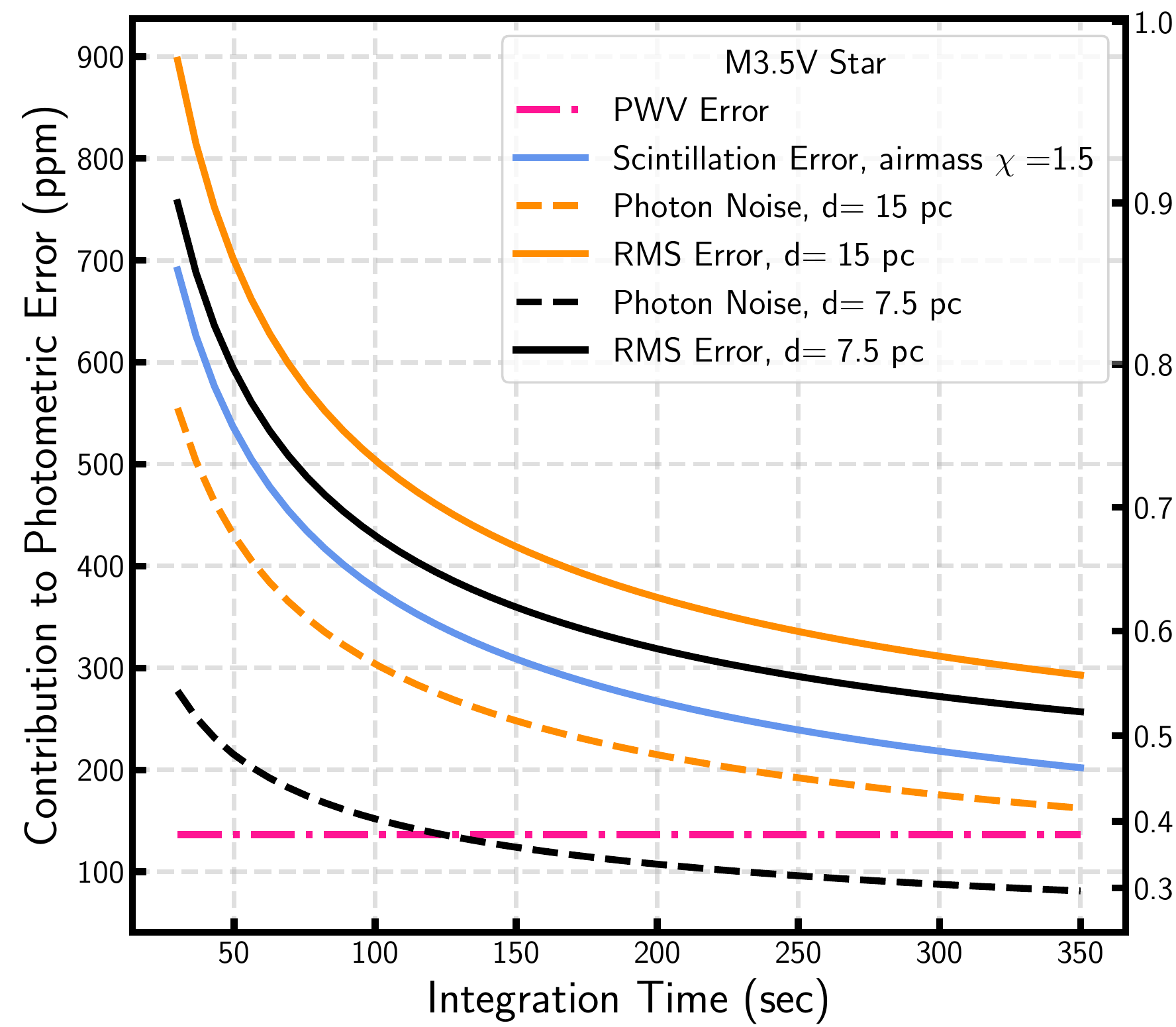}
\includegraphics[height=6.82cm]{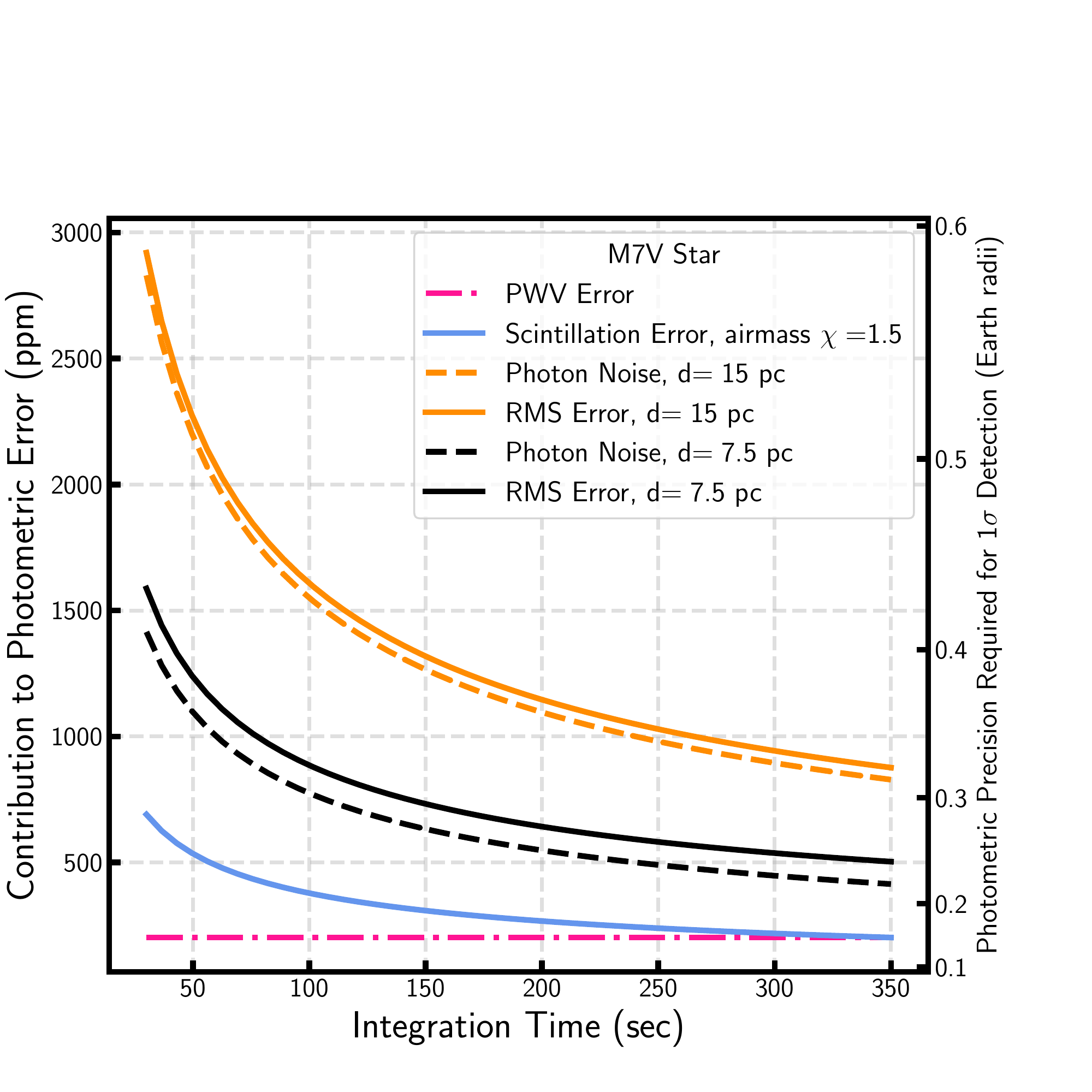}
\end{tabular}
\end{center}
\caption[example] 
{\footnotesize \label{fig:performance} Contribution of the PWV error, scintillation error, and photon noise to the \tierras RMS photometric precision, $\sigma_{\rm tot}$ (Equation \ref{eq:toterr}) when observing an M3.5V (left) and an M7V (right) target star, placed either 15 pc (orange curves) or 7.5 pc (black curves) away from the Earth. The PWV error through the \tierras filter is plotted in pink (dash-dotted), assuming a PWV range of 0-12 mm. The predicted scintillation error is plotted in blue (solid), and assumes that both stars are observed at an airmass of $\chi = 1.5$. The target photon noise for both stars is shown in orange and black (dashed) when placed at 15 pc and 7.5 pc, respectively. Similarly, the RMS noise for both stars is shown in orange and black (solid) when placed at 15 pc and 7.5 pc, respectively. The y-axis on the left side of both plots quantifies error contribution in ppm. The y-axis on the right side of both plots quantifies error in units of Earth radii. Specifically, it is given by $\sqrt{\sigma_{\rm tot}} \times R_*$, where $R_*$ is the estimated stellar radius. We assume a stellar radius of $0.3 R_{\odot}$ for the M3.5V star, and of $0.1 R_{\odot}$ for the M7V star \cite{Nu08}.}
\end{figure}

Figure \ref{fig:performance} shows our results for both M dwarf stars at 15 pc and 7.5 pc. Our PWV error is given by 136 ppm for the M3.5V star, and 202 ppm for the M7Ve star, fulfilling our design requirement for this systematic error. In the case of the mid M dwarf star (Gliese 555) at 15 pc, we must integrate for 304 seconds to achieve a photometric precision equal to a $3\sigma$ detection of an Earth transit (310 ppm). If the same target were 7.5 pc away, our expected photometric precision would be 272 ppm for a 5 min integration. In the case of the late M dwarf (vb8) at 15 pc, we must integrate for about 32 seconds to reach a photometric precision equal to an Earth transit (2795 ppm), and can reach 947 ppm after integrating on the star for 5 minutes. If vb8 were 7.5 pc away, we could reach a photometric precision as low as 539 ppm for a 5 min integration.

\subsection{Future Work and Observatory Upgrades}\label{subsec:future}

In tandem with in-lab camera testing and optical assembly, we are building a set of telescope baffles to minimize scattered light at the image plane, and exploring the design of a custom-made nano-fabricated beam-shaping diffuser to mold the point spread function into a broad top-hat shape. Below, we provide a quick overview of work being done on these fronts:

\textit{ Telescope Baffling --} Despite having an open tube design, the 1.3-m telescope was not equipped with light baffles while the 2MASS and PAIRITEL projects were carried out, seeing as both projects operated in the NIR. In contrast, baffling is crucial towards fulfilling our photometric precision goals. We designed telescope baffles for the \tierras telescope following the iterative graphical optimization in Ref. \citenum{Pr68}. Our design prevents sky light and dome light from reaching the image plane directly, or after a single scatter off a baffle surface. It also guarantees that light rays with semi-field angles $u$ below the maximum allowable semi-field angle, $u_{pr}$, reach the image/focal plane, but only after reflecting off the primary and secondary mirrors. Considering our frame-transfer CCD architecture (Figure \ref{fig:ccdqe}) we defined $u_{pr} = 0.28 \arcdeg$, which allows sky light to fill the effective imaging area of our detector, while also preventing vignetting. While baffling inevitably results in additional primary mirror obstruction, our design increases the photon noise contribution to the total photometric error by $<5\%$. We are currently finalizing the baffle design and preparing for building them in house. 

\textit{Diffuser Design --} We are considering using a custom-made nano-fabricated beam-shaping diffuser to mold the point spread function into a broad and stable top-hat shape \cite{Ra04}. By spreading the light over more pixels, we can increase the dynamic range of our observations, while becoming immune to seeing variations and small pointing drifts. Ref. \citenum{St17} have employed this technology to demonstrate on-sky photometric precision of 137 ppm in the NIR (in 30 min bins). Given that diffusers work optimally in near-collimated beams, and that the focal reducer design does not allow us to fit any additional components between lenses 1 through 4 and the CCD (Figure \ref{fig:optics}), a $6$ mm thick diffuser would have to be at least as large as the filter, and placed a minimum distance $D_{\rm diff} = 189$ mm away from the image plane. The approximate FWHM of a diffused stellar spot on the image plane is given by $D_{\rm diff} \tan \theta$, where $\theta$ is the opening angle of the diffuser; its most important design parameter. It follows that spreading a stellar spot radially over 10-20 ($15 \, \mu$m size) pixels in the \tierras image plane would require diffusion angles $\theta$ of 0.046-0.091$^{\circ}$. The narrower the diffusion angle, the greater the fabrication challenges for diffuser manufacturers, which, in tandem with the required diameter of the diffuser, calls for a custom-made and therefore expensive optical element. Due to the 1.3-m telescope architecture, the diffuser would be permanently located in the light path, requiring that comparison stars in all the fields of interest be about $9 \arcsec$ away from the target star in order to prevent blending. For all these reasons, we intend to carry out a more detailed cost-benefit analysis after first light in order to determine whether a diffuser would significantly increase the \tierras scientific yield. 

\section{Summary}\label{sec:summary}

We have provided an overview of the current status of the \tierras Observatory, a now fully refurbished 1.3-m telescope atop the ridge of Mt. Hopkins, AZ. By means of a custom filter, \tierras will limit the PWV error to 136 ppm for M3.5V stars, and 202 ppm for the M7V star, fulfilling our design requirement. For an M3.5V star, we predict that a 5 min integration will be enough to limit stochastic errors below the precision required to detect an Earth transit with $3\sigma$ significance (310 ppm). In the case of an M7V star at 15 pc, we must only integrate for 32 seconds to reach a photometric precision equal to an Earth transit with $3\sigma$ significance (2795 ppm), and can reach 947 ppm after integrating on the star for 5 minutes. We are in the process of assembling the \tierras filter, as well as the focal reducer and field flattener in the lab. Testing of the CCD camera in its custom dewar is also underway. The \tierras Observatory will have an early 2021 first light, followed by several months of commissioning, software development and robotization before beginning science operations in the spring of 2021.  

\acknowledgments 

The authors would like to thank the staff at the F. L. Whipple Observatory atop Mt. Hopkins for their assistance in the refurbishment and maintenance of the 1.3-m telescope. We also thank Nelson Caldwell for his guidance on the design of the \tierras baffles. The \tierras Observatory project is supported by a grant from the John Templeton Foundation and a grant from the Harvard Origins of Life Initiative. The opinions expressed in this publication are those of the authors and do not necessarily reflect the views of the John Templeton Foundation. JGM is supported by NSF Graduate Research Fellowship Grant No. DGE1745303, and a Ford Foundation Predoctoral Fellowship administered by the National Academies of Sciences, Engineering, and Medicine. 


\bibliography{report} 
\bibliographystyle{spiebib} 

\end{document}